\documentclass[preprintnumbers,aps,11pt,floatfix,nofootinbib]{revtex4} 
\long\def\del #1 \enddel { }

\usepackage{graphicx}
\usepackage{amsmath}
\usepackage{amssymb}
\usepackage{subfigure}
\usepackage{epsfig}
\usepackage{psfig}

\newdimen\arrayruleHwidth
\makeatletter
\def\Hline{\noalign{\ifnum0=‘}\fi\hrule \@height \arrayruleHwidth
\futurelet \@tempa\@xhline}

\makeatother

\begin{document}

%%%%%%%%%%%% Definitionen %%%%%%%%%%

\def\beq{\begin{equation}}
\def\eeq{\end{equation}}

\def\bea{\arraycolsep .1em \begin{eqnarray}}
\def\eea{\end{eqnarray}}
\def\Tr{{\rm Tr}}

\def\dq{{q \llap{/}}\,}
\def\dk{{k \llap{/}}\,}
\def\step{\\[-2ex]}
\def\bigstep{\\[3ex]}

\let\de=\delta
\let\Ga=\Gamma
\let\eps=\epsilon
\let\om=\omega
\def\bpi{\mbox{\boldmath$\pi$}}
\def\bphi{\mbox{\boldmath$\phi$}}

\def\Eq#1{Eq.~(\ref{#1})}
\def\eq#1{(\ref{#1})}
\def\fig#1{fig.~(\ref{#1})}

\def\s0#1#2{\mbox{\small{$ \frac{#1}{#2} $}}}
\def\0#1#2{\frac{#1}{#2}}

\def\llangle{\left\langle}
\def\rrangle{\right\rangle}

\begin{center}

\thispagestyle{empty}

{\normalsize\begin{flushright}CERN-TH/2007-037 \\[12ex] 
\end{flushright}}

\mbox{\large \bf Towards Functional Flows for Hierarchical Models}
\\[6ex]

{Daniel F. Litim}\footnote{D.Litim@sussex.ac.uk, Daniel.Litim@cern.ch}
\\[3ex]
{\it
Department of Physcis and Astronomy\\ 
University of Sussex, Brighton, BN1 9QH, U.K.}
\\[3ex]
{\it Theory Group, Physics Division\\ 
CERN, CH-1211 Geneva 23.}%\\[1ex]
\\[10ex]

{\small \bf Abstract}\\[2ex]
\begin{minipage}{14cm}{\small 
    The recursion relations of hierarchical models are studied and contrasted
    with functional renormalisation group equations in corresponding
    approximations.  The formalisms are compared quantitatively for the Ising
    universality class, where the spectrum of universal eigenvalues at
    criticality is studied. A significant correlation amongst scaling
    exponents is pointed out and analysed in view of an underlying
    optimisation.  Continuous functional flows are provided which match with
    high accuracy all known scaling exponents from Dyson's hierarchical model
    for discrete block-spin transformations.  Implications of the results are
    discussed.}
\end{minipage}
\end{center}

%PACS: 05.10.Cc, 11.10.Hi, 11.10.Kk, 64.60.Fr
\newpage
\pagestyle{plain}
\setcounter{page}{1}
\noindent 
{\bf 1. Introduction}\\[-1ex]

Renormalisation group methods \cite{Zinn-Justin:1989mi}, and in particular
Wilson's renormalisation group \cite{Wilson:1973jj}, 
play an important role in the study of physical systems at strong coupling
and/or large correlations lengths.  Differential
implementations of Wilson's idea  
\cite{Tetradis:1995br,Morris:1998da,Litim:1998yn,Litim:1998nf,Bagnuls:2000ae,%
  Berges:2000ew,Polonyi:2001se,Salmhofer:2001tr} rely on an
appropriately-introduced momentum cutoff leading to flow equations for running
couplings and $N$-point functions, which can be studied with a large variety
of analytical and numerical methods.  Numerical stability and reliability in
the results is ensured through powerful control and optimisation techniques
\cite{Litim:2000ci,Litim:2001up,Litim:2001fd,Pawlowski:2005xe}.  A different
implementation of Wilson's idea is realised in hierarchical models of lattice
scalar theories \cite{Dyson:1968up,Wilson:1971dh,%
  Golner:1973,Baker:1972,Meurice:2007zg}. Hierarchical renormalisation group
transformations are often discrete rather than continuous.  Here,
sophisticated numerical methods have been developed to extract the relevant
physics, most notably for high-accuracy studies of scaling exponents for
scalar models at criticality \cite{KochWittwer:1988,Pinn:1994st,Godina:1998uz}
and related theories (see \cite{Meurice:2007zg} and references therein).
\step

Given the close similarity of the underlying principles, it is natural to ask
whether Wilsonian (functional) flows can be linked explicitly, and on a
fundamental level, to hierarchical models.  If so, this link would provide a
number of benefits. It will make powerful functional and numerical methods
available to the study of hierarchical models. Vice versa, the numerical tools
for hierarchical models could be employed for functional flows in specific
approximations.  Furthermore, an explicit link may lead to a path integral
representation of hierarchical models, allowing for systematic improvements
beyond a standard kinetic term.  Finally, well-developed optimisation
techniques for functional flows could be taken over for hierarchical models as
well.  \step

In the limit of continuous hierarchical block-spin transformations, an
explicit link between Dyson's hierarchical model
\cite{Dyson:1968up,Baker:1972} and the Wilson-Polchinski flow
\cite{Polchinski:1983gv} in the local potential approximation has been
established long ago by Felder \cite{Felder:1987}.  Following a conjecture of
\cite{Litim:2001fd}, this link has been extended
\cite{Litim:2005us,Morris:2005ck} to include optimised versions
\cite{Litim:2000ci,Litim:2001up} of Wetterich's flow for the effective average
action \cite{continuum}. These interrelations have recently been backed-up by
extensive numerical studies of critical potentials and scaling exponents to
high accuracy from either formalism \cite{Bervillier:2007rc}.  \step

In this paper, we evaluate the more general case and ask whether hierarchical
models for discrete block-spin transformations are linked to functional flows
with continuous renormalisation group transformations.  We first contrast the
basic setups for functional flows (Sec.~2), background field flows (Sec.~3),
and hierarchical models (Sec.~4).  At a Wilson-Fisher fixed point, underlying
similarities and differences are worked out and compared for the leading
scaling exponent (Sec.~5). An extensive numerical study of the eigenvalue
spectrum of the Ising universality class from functional flows is performed
(Sec.~6). A strong correlation of scaling exponents is established and
analysed (Sec.~7).  It is shown that specific functional flows match the
leading and subleading scaling exponent from Dyson's hierarchical models for
discrete transformation parameter to high accuracy (Sec.~8).  We close with a
discussion of the results and further implications (Sec.~9).\bigstep

%********|*********|*********|*********|*********|*********|*********|****
\noindent 
{\bf 2. Functional Flows}\\[-1ex]%\step
%********|*********|*********|*********|*********|*********|*********|****

Wilsonian (functional) flows integrate-out quantum fluctuations within a path
integral representation of quantum field theory. In their simplest form, they
are generated through a cutoff term quadratic in the field added to the
Schwinger functional, where the (classical) action is replaced by $S\to
S+\Delta S_k$ and $\Delta S_k\sim \int dq \phi(q) R_k(q^2)\phi(-q)$. The
infrared momentum cutoff $R_k(q^2)$ ensures that the propagation of small
momentum modes $q^2\ll k^2$ is suppressed, while the large momentum modes
$q^2\gg k^2$ remain unaffected. Under an infinitesimal change in the Wilsonian
(infrared) cutoff scale $k$, the effective action $\Gamma_k$ changes according
to its functional flow, which reads ($t=\ln k$)
\begin{equation}
\partial_t\Gamma_k=\012\Tr\, \left(\Gamma_k^{(2)}+R_k\right)^{-1}\,
\partial_t R_k
\label{ERG}
\end{equation}
in the form put forward by Wetterich \cite{continuum}. The trace denotes a
momentum integration and a summation over fields.  The factor $\partial_t R_k$
in the integrand is peaked in the vicinity of $q^2\approx k^2$. The cutoff
function $R_k$ obeys $R_k(q^2)\to 0$ as $k^2/q^2\to 0$, $R_k(q^2) > 0$ as
$q^2/k^2\to 0$, and $R_k(q^2)\to \infty$ as $k\to \Lambda$, and can be chosen
freely elsewise, $e.g.$ \cite{Litim:2000ci}.  It ensures that the flow is
well-defined, thereby interpolating between an initial action $S$ at
$k=\Lambda$ in the ultra\-violet (UV) and the full quantum effective action
$\Gamma\equiv\Gamma_{k=0}$ in the infrared $k\to 0$. \step

In addition to providing a momentum cutoff, the function $R_k(q^2)$ also
controls the stability and convergence of subsequent expansions
\cite{Litim:2000ci,Litim:2001up,Litim:2001fd,Litim:2001dt}.  Therefore, it is
possible to identify optimised momentum cutoffs -- within given systematic
expansions -- which improve the physical result
\cite{Litim:2000ci,Litim:2001up,Litim:2002cf}. The construction of optimised
cutoffs \cite{Litim:2000ci,Litim:2001up,Litim:2001fd,Pawlowski:2005xe} is
central to extract reliable results also in more complex theories including
$e.g.$~QCD \cite{Pawlowski:2003hq}, quantum gravity \cite{Lauscher:2001rz},
thermal physics \cite{Litim:1998yn,Litim:1998nf, Litim:2006ag}
and critical phenomena
\cite{Litim:2002cf,Blaizot:2004qa,Bervillier:2007rc}.\step

Below, we are interested in $3d$ scalar theories at criticality, where we can
sent the ultraviolet scale $\Lambda\to\infty$. To leading order in the
derivative expansion, the effective action reads $\Gamma_k=\int
d^3x[\s012\partial_\mu\phi\partial_\mu\phi+U_k(\bar\rho)]$ and
$\bar\rho=\s012\phi^2$. Introducing $r(y)=R_k(q^2)/q^2$ with $y=q^2/k^2$, we
find
\begin{equation}
\partial_t u=-3u+\rho u' 
+ \int_0^\infty dy \frac{-y^{3/2}\,r'(y)}{y(1+r)+u'+2\rho u''}
\label{flowERG}
\end{equation}
with $u(\rho)=U_k(\bar\rho)/k^3$ and $\rho=\bar\rho/k$.  An irrelevant
constant originating from the angular integration has been rescaled into the
potential and the fields. For the optimal cutoff $R_{\rm
  opt}=(k^2-q^2)\theta(k^2-q^2)$ with $r_{\rm opt}=(1/y-1)\theta(1-y)$, the
flow reads \cite{Litim:2001up}
\begin{equation}
\partial_t u=-3u+\rho u' + \frac{1}{1+u'+2\rho u''}
\label{flowOpt}
\end{equation}
after an additional rescaling. This flow is integrated analytically in the
limit of a large number of scalar fields \cite{Litim:1995ex}. We note that the
universal content of the flow \eq{flowOpt} is equivalent to the
Wilson-Polchinski flow in the local potential approximation
\cite{Litim:2001fd,Litim:2005us,Morris:2005ck}.  \bigstep

%********|*********|*********|*********|*********|*********|*********|****
\noindent 
{\bf 3. Background Field Flows}\\[-1ex]%\step
%********|*********|*********|*********|*********|*********|*********|****

A different form of the flow \eq{ERG} is obtained for momentum cutoffs which
depend additionally on a background field $\bar\phi$.  Background fields are
most commonly used for the study of gauge theories \cite{Reuter:1993kw}, see
\cite{Reuter:1997gx,Litim:2002ce,Lauscher:2001rz} for applications. They have
also been employed for a path integral derivation of (generalised) proper-time
flows \cite{Litim:2002hj,Litim:2002xm}.\step 

In the presence of background fields, the functional $\Gamma_k[\phi]$ turns
into a functional of both fields, $\Gamma_k[\phi,\bar\phi]$.  In order to
maintain the one-loop exactness of \eq{ERG}, the momentum cutoff can only
depend on the background field, but not on the propagating field.  Following
\cite{Litim:2002hj}, we introduce $x=\Gamma^{(2,0)}[\phi,\phi]$ and $\bar
x=x[\phi=\bar\phi]$, where $\Gamma_k^{(n,m)}[\phi,\bar\phi]\equiv \delta^n
\delta^m\Gamma_k/\delta\phi^n\delta\bar\phi^m$. We chose momentum cutoffs of
the form $R_k(q^2)\to \bar x\, r[\bar x]$, which depend now on the background
field.  Here, the regulator cuts off both large momentum modes $q^2\gg k^2$
and large field amplitudes with $\Gamma_k^{(2,0)}\gg k^2$.  The full advantage
of background fields becomes visible once they are identified with the
physical mean, leading to the functional
$\Gamma_k[\phi,\bar\phi=\phi]\to\Gamma_k[\phi]$.  The resulting flow is closed
provided $\Gamma_k^{(2)}[\phi]=\Gamma_k^{(2,0)}[\phi,\phi]$. For scalars, this
relation becomes exact in the infrared limit studied below (for gauge fields,
see \cite{Litim:2002ce}). Using the momentum cutoffs \cite{Litim:2002hj}
\begin{eqnarray}\nonumber
r_{{\rm PT},m}[x]= 
\exp\left(\s01m\left(\s0{mk^2}{x}\right)^m\,{}_2 F_1[m,m;m+1;-\s0{m\, 
k^2}{x}]\right)-1\,, 
\end{eqnarray}
we are lead to the background field flow
\begin{eqnarray}\label{flowm}
\partial_t \Gamma_k = 
\Tr\, \left({k^2 \over k^2+x/m}\right)^m 
+
{1\over 2} \Tr\left[\, {r_{{\rm PT},m} \over x(1+r_{{\rm PT},m})}
-\left({k^2 \over k^2+x/m}\right)^m {1\over x}\right]
\,  \partial_t x\,. 
\end{eqnarray}
If the term $\sim \partial_tx$ on the right hand side is dropped -- meaning
that additional flow terms originating from the implicit scale dependence in
the momentum cutoff are neglected over the leading term -- the flow \eq{flowm}
reduces to the proper-time flow of Liao \cite{Liao:1997nm}. A general
proper-time flow is a linear combination of the first term in \eq{flowm} for
various $m$ \cite{Litim:2002xm}; see \cite{Bonanno:2000yp,Schaefer:1999em} for
applications.\step

Next, we specialise to the proper-time approximation to leading order in the
derivative expansion. The flow equation for the effective potential takes the
very simple form
\begin{equation}
\partial_t u=-3u+\rho u' 
+ \frac{1}{(m+u'+2\rho u'')^{m-3/2}}\,,
\label{flowPT}
\end{equation}
where $m$ parameterises the momentum cutoff, and an irrelevant constant factor
has been rescaled into the potential and the fields. For $m\in
[1,\s052]$, the flow \eq{flowPT} is mapped onto the flow \eq{flowERG}
\cite{Litim:2002hj}.  At $m=\052$, the flow \eq{flowPT} is equivalent to
\eq{flowOpt}, modulo a trivial rescaling. As a final remark, we note that this
proper-time flow is also obtained from linear combinations of higher
scale-derivatives of Callan-Szymanzik flows, without relying on background
fields \cite{Litim:2002xm}. In this representation, the approximation leading
to \eq{flowPT} consists in the neglection of higher order flow terms
$\sim\partial_t^n\Gamma_k^{(2)}$.\bigstep
 
%********|*********|*********|*********|*********|*********|*********|****
\noindent 
{\bf 4. Hierarchical Models}\\[-1ex]%\step
%********|*********|*********|*********|*********|*********|*********|****

Several hierarchical models for an effective potential $v(\varphi)$ of a
lattice scalar field have been introduced in the literature
\cite{Dyson:1968up,Wilson:1971dh,Baker:1972} (see also \cite{Meurice:2007zg}).
The hierarchical transformation laws relate the potential $v(\varphi)$ at
momentum scale $k/\ell$ with an average in field space over $v(\varphi)$ at
momentum scale $k$, where $\ell\ge 1$ is the renormalisation group step
parameter.  We restrict ourselves to the three-dimensional case; the
generalisation to arbitrary dimensions is straightforward.\step

In Dyson's model \cite{Dyson:1968up,Baker:1972}, the renormalisation group
step $k\to k/\ell$ for the potential is expressed as
\begin{equation}\label{DHM}
e^{-v_{k/\ell}(\varphi)}=
\int_{-\infty}^{+\infty}\, d\xi\, \mu_\ell(\xi)
\,e^{-\ell^3\, v_k(\ell^{-1/2}\varphi+\xi)}
\end{equation}
The details of the averaging procedure are encoded in the measure factor
$\mu_\ell(\xi)$, in the $\xi$-dependence of the potential on the right-hand
side of \eq{DHM}, and in the choice for the decimation parameter $\ell$.  As
is evident from \eq{DHM}, a decimation parameter $\ell=2^{1/3}$ -- employed
for most numerical studies
\cite{KochWittwer:1988,Pinn:1994st,Godina:1998uz,Meurice:2007zg}
-- corresponds to a volume decimation of $\ell^3=2$ at each iteration. For
Dyson's model, the measure is chosen as
$\mu_\ell(\xi)=(\pi\,\sigma(\ell))^{-1/2}\exp(-\xi^2/\sigma(\ell))$
\cite{Meurice:2007zg}, where we require $\sigma(\ell)> 0$ for $\ell\neq 1$,
and $\sigma(1)=0$ with $\sigma'(1)\neq 0$. A standard choice is
$\sigma(\ell)=2(\ell-1)$ \cite{Felder:1987}.  By definition, \eq{DHM}
describes a flow towards the infrared for decimation parameters $\ell\ge 1$.
For $\ell\to 1$, the hierarchical transformation \eq{DHM} becomes continuous
and the measure factor turns into a $\delta$-function $\mu_{\ell\to
  1}(\xi)\to\delta(\xi)$.  Performing $-\ell\partial_\ell$\eq{DHM}, which is
equivalent to $k\partial_k$\eq{DHM}, we arrive at a differential flow equation
for the effective potential \cite{Felder:1987}
\begin{equation}
\partial_t v=-3v+\s012\varphi v' - v''+ (v')^2\,,
\label{flowWP}
\end{equation}
where an irrelevant factor is rescaled into the fields and the potential;
$t=\ln k$. The interaction terms in \eq{flowWP} originate from the
scale-derivative of the measure $-\ell\partial_\ell\, \mu_\ell(\xi)$, which
reads $\s014\,\sigma'(1)\,\delta''(\xi)$ in the limit $\ell\to 1$.  This
highlights the relevance of the measure factor in hierarchical models. Our
normalisation corresponds to the choice $\sigma'(1)=4$ to match with
\cite{Bervillier:2007rc}.  The limit \eq{flowWP} is independent of $\sigma$,
but at $\ell\neq 1$, we expect that scaling solutions and exponents from
\eq{DHM} depend on it.  \Eq{flowWP} is the well-known Wilson-Polchinski flow
\cite{Wilson:1973jj,Polchinski:1983gv}. We therefore conclude that the
potential in \eq{flowWP} is related to the potential in \eq{flowOpt} by a
Legendre transformation \cite{Morris:2005ck,Bervillier:2007rc}.\step

A different version of a hierarchical model has been introduced by Wilson
\cite{Wilson:1971dh}. Here, the recursion relation is written as
\begin{equation}\label{WHM}
e^{-v_{k/\ell}(\varphi)}=
\int_{-\infty}^{+\infty}\, d\xi\, \mu_\ell(\xi)\,
e^{-\s012\ell^3\,[ v_k(\ell^{-1/2}\varphi+\xi)
+ v_k(\ell^{-1/2}\varphi-\xi)]}\,.
\end{equation}
In Wilson's original model, the $\xi$-dependence of the measure is
$\mu_\ell(\xi)=N_\ell\,\exp(-\xi^2)$, where the normalisation factor $N_\ell$
is $\xi$-independent \cite{Meurice:2007zg}.  The measure factor is different
from the one in Dyson's model, because the Gaussian width is
$\ell$-independent. If instead we employ the measure of Dyson's model, the
limit $\ell\to 1$ can be performed analytically.\footnote{\label{Rescaling}
  The variance of the Gaussian measure in \eq{WHM} can be changed by an
  explicit rescaling of the fields as $\varphi\to\varphi/\sqrt{\sigma}$ for
  finite $\sigma$, see~\cite{Meurice:2007zg}.  Rescaling also the integration
  variable $\xi\to\xi/\sqrt{\sigma}$, and denoting the potential in terms of
  the rescaled fields again as $v(\varphi)$, we obtain \eq{WHM} with a
  rescaled measure $\mu_\ell(\xi)=N_\ell\,\sigma^{-1/2}\exp(-\xi^2/\sigma)$.
  It agrees with the measure of Dyson's model for $N_\ell=\pi^{-1/2}$ and
  $\sigma=\sigma(\ell)$.  I thank Y.~Meurice for e-mail correspondence on this
  point.} Up to a trivial rescaling, we find
\begin{equation}
\partial_t v=-3v+\s012\varphi v' - v''\,.
\label{flowWHM}
\end{equation}
In contrast to the Wilson-Polchinski flow \eq{flowWP}, the non-linear term
$(v')^2$ is absent. This comes about because the integrand of \eq{WHM} -- as
opposed to the integrand of \eq{DHM} -- is manifestly symmetric under
$\xi\to-\xi$.  Numerical evaluations of \eq{WHM} have been reported
in~\cite{Wilson:1971dh,Meurice:1996bh}. For other representations of
hierarchical models we refer to \cite{Meurice:2007zg} and references therein.
\bigstep

%********|*********|*********|*********|*********|*********|*********|****
\noindent 
{\bf 5. Matching Hierarchical Models}\\[-1ex]%\step
%********|*********|*********|*********|*********|*********|*********|****

In order to match hierarchical models by functional flows, we have to detail
the scheme dependences of physical observables in either formalism. In the
functional RG framework, the fully integrated flow is independent of the
momentum cutoff $R_k(q^2)$ chosen for the integration.  Scheme dependences,
which enter as a consequence of truncations of $\Gamma_k[\phi]$, have been
discussed extensively in the literature
\cite{Ball:1995ji,Litim:2000ci,Litim:2001up,Litim:2001fd,Litim:2001dt}.  Their
origin is easily understood. Since the momentum cutoff $R$ in \eq{ERG} couples
to all operators in the theory, the missing back-coupling of operators
neglected in a given truncation can result in a spurious dependence of
physical observables evaluated either from $\Gamma_0[\phi]$, or from a fixed
point solution $\Gamma_*[\phi]$. The scheme dependence is reduced by
identifying those momentum cutoffs, which, in a given truncation, lead to an
improved convergence and stability of the flow. \step

\begin{figure}
\begin{center}
\unitlength0.001\hsize
\begin{picture}(700,500)
\put(70,480){\Large $\nu$}
\put(410,450){Callan-Symanzik}
\put(455,400){sharp cutoff}
\put(500,210){WP/opt}
\put(475,60){mean field}
\put(130,170){\small exact RG}
\put(140,100){\small proper-time RG}
\put(290,27){\small Wilson HM}
\put(370,170){\small Dyson HM}
\includegraphics[width=.6\hsize]{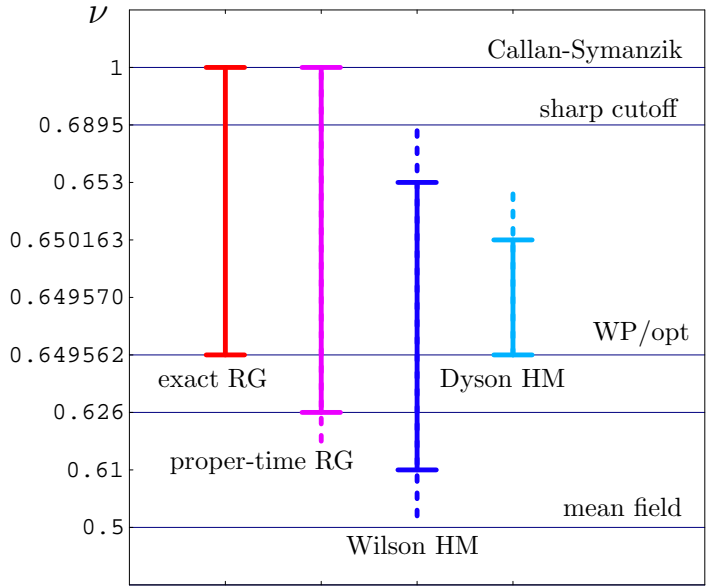}
\end{picture}
\vskip.2cm
\caption{\label{fHMvsFRG} 
  (colour online) Comparison of scaling exponent $\nu$ from different
  functional flows (RG) in the local potential approximation, and hierarchical
  models (HM).  The solid lines indicate the range of values obtained in the
  literature. The dashed lines, if present, indicate that the underlying
  parameter space has not been exhausted. The horizontal lines from top to
  bottom indicate the results for Callan-Symanzik flows, the sharp cutoff
  flow, the Wilson-Polchinski (optimal) flow, and the mean field result. A
  non-linear rescaling of the $\nu$-axis is introduced for display purposes
  only (see main text).  Colour coding: exact RG (red), exact background field
  RG in the proper-time approximation (violet), Wilson's hierarchical model
  (blue) and Dyson's hierarchical model (light blue).}\vskip-.5cm
\end{center}
\end{figure}

In Fig.~\ref{fHMvsFRG}, we discuss the scheme dependence quantitatively for
the leading order scaling exponent $\nu$ at a fixed point of the $3d$ Ising
universality class \cite{Litim:2001dt}.  Within exact flows \eq{flowERG}, the
full $R_k$-dependence has been studied in \cite{Litim:2002cf} by evaluating
the fixed points of \eq{flowERG} for general cutoffs (Fig.~\ref{fHMvsFRG},
first column). The main result is that the range of achievable values is
bounded from above and from below. The upper bound is attained for
Callan-Symanzik type flows with $R_k\sim k^2$. The lower bound with
$\nu=\nu_{\rm opt}$ is attained with the optimal flow \eq{flowOpt}, and
hence equivalent to the Wilson-Polchinski flow. The sharp cutoff result is
indicated for comparison.\step

%*************************************************************************
% Figure 1 - scaling exponent nu
%*************************************************************************
%*************************************************************************

The proper-time flow \eq{flowPT} rests on an intrinsically different
truncation, because implicit dependences on the background field have been
neglected as well as higher order flow terms proportional to the flow of
$\Gamma^{(2)}$; see Sec.~3.  Therefore scheme dependences are quantitatively
different. In the approximation \eq{flowm}, the $m$-dependence of scaling
exponents from \eq{flowPT} has been studied in
\cite{Bonanno:2000yp,Litim:2002hj} (Fig.~\ref{fHMvsFRG}, second column).  The
range of values is again bounded from above by a Callan-Symanzik flow.  The
lower bound is achieved for $m\to\infty$. We note that the range of values
exceeds those achievable within (standard) exact flows. The lower bound may be
overcome once the additional flow terms, neglected here, are taken into
account \cite{Litim:2002hj}. This is indicated by the dashed line. \step

Next we consider scheme dependences of hierarchical models. Based on their
construction, we expect that physical observables depend on the averaging
procedure, on the measure factor $\mu_\ell$, and on the decimation parameter
$\ell$. It has proven difficult to systematically include wave function
renormalisations and higher order operators in hierarchical models, and it is
therefore not known whether the scheme dependence vanishes upon higher order
corrections \cite{Golner:1973,Meurice:2007zg}.  Still, the scheme dependence
should give a reasonable estimate for the underlying error in the model
assumptions, in particular in comparison with functional methods.  \step

The $\ell$-dependence of Wilson's model \eq{WHM}, originally constructed for
$\ell\approx 2$, has been studied in \cite{Meurice:1996bh} in the range
$\ell\in[2^{1/3},2]$ (Fig.~\ref{fHMvsFRG}, third column). The full line covers
the range of values obtained in the literature, while the dashed lines
indicate that the underlying parameter space has not been exhausted. Because
of \eq{flowWHM} being linear in the potential as opposed to \eq{flowWP}, we
expect a strong $\ell$-dependence, possibly a discontinuity, in the limit
$\ell\to 1$. The slope $\ell\partial_\ell\, \nu(\ell)$ along the data points
with $\ell>2^{1/3}$ is negative, meaning that $\nu(\ell)$ increases for
smaller $\ell$. We stress that Wilson's HM has an overlap both with exact
flows and proper-time flows.  Therefore, it is possible to map $\nu(\ell)$ of
Wilson's model for certain decimation parameters $\ell$ onto $\nu(R)$ from
functional flows with appropriately chosen $R$.  On the other hand, for some
decimation parameter $\ell$, Wilson's model can only be mapped onto
proper-time flows but not on exact flows, while for some decimation parameter
it cannot be mapped onto either of them. \step

The $\ell$-dependence of Dyson's model \eq{DHM} is displayed in
Fig.~\ref{fHMvsFRG}, fourth column.  The full line connects the known results
at $\ell=1$ \cite{Litim:2002cf,Bervillier:2007rc}, $\ell=2^{1/3}$
\cite{KochWittwer:1988,Godina:1998uz}
and $\ell=2$ \cite{KochWittwer:1988}. The dashed line towards larger values
for $\nu$ indicates that the parameter space $\ell\ge 1$ has not been
exhausted. We note that the $\ell$-dependence is very weak, with a tiny slope
in the range of $\ell$-values covered.  The important observation is that the
slope $\ell\partial_\ell\, \nu(\ell)$ is positive in the vicinity of
$\ell\approx 1-2$, implying that $\nu(\ell)>\nu(1)$ for $\ell>1$.
Consequently, it is possible to map the scaling exponent $\nu(\ell)$ at
discrete block-spin transformation $\ell>1$ onto $\nu(R)$ from functional
flows for specific momentum cutoff $R$, both within the standard exact flows
and within proper-time flows. This supports the conjecture that Dyson's model
can be mapped onto functional flows.\bigstep

%********|*********|*********|*********|*********|*********|*********|****
\noindent
{\bf 6. Spectrum of Eigenvalues}\\[-1ex]%\step
%********|*********|*********|*********|*********|*********|*********|****

Whether the observations of the preceeding section can be promoted to a full
map between the formalisms crucially depends on further observables including
the subleading scaling exponents.  Here and in the following section, we study
the spectrum of universal eigenvalues (scaling exponents) form functional
flows \eq{flowERG} to high accuracy.  A fixed point solution $u_*\neq const.$
of \eq{flowERG} is characterised by the universal eigenvalues of
eigenperturbations in its vicinity.  We denote the ordered set of eigenvalues
as ${\cal O}(R)=\{\omega_i(R),\, i=0,\cdots,\infty\}$, with
$\omega_i<\omega_j$ for $i<j$.\footnote{In our conventions, the sole negative
  eigenvalue at the Wilson-Fisher fixed point is $\omega_0$.}  In addition to
the leading exponent $\nu(R)\equiv-1/\omega_0$, we study the first three
subleading scaling exponents $\omega(R)\equiv \omega_1(R)$, $\omega_2(R)$ and
$\omega_3(R)$ within the exact flow \eq{flowERG} for various cutoffs and
coarse graining parameters.  \step

For the numerical analysis, we introduce several classes of momentum cutoffs
defined through $r_{\rm mexp}=b/((b+1)^{y}-1)$, $r_{\rm exp}=1/(\exp c y^b
-1)$; $r_{\rm mod}=1/(\exp[c(y+(b-1)y^b)/b] -1)$, with $c=\ln 2$; and $r_{{\rm
    opt},n}=b(1/y-1)^n\theta(1-y)$.  These cutoffs include the sharp cutoff
$(b\to\infty)$ and asymptotically smooth Callan-Symanzik type cutoffs $R_k\sim
k^2$ as limiting cases. The larger the parameter $b$, for each class, the
`sharper' the corresponding momentum cutoff. The cutoff $r_{{\rm opt},n}$
probes a two-dimensional parameter space in the vicinity of $r_{\rm opt}$ to
which it reduces for $b=1$ and $n=1$.  For integer $n$, $r_{{\rm opt},n}$ is a
$C^{(n+1)}$ function.  In addition, we consider the cutoffs $r_{\rm
  mix}=\exp[-b(\sqrt{y}-1/\sqrt{y})]$ and $r_{\rm
  mix,opt}=\exp[-\s01b(y^b-y^{-b})]$, which obey $r_{\rm mix}(1/y)=1/r_{\rm
  mix}(y)$. Note that we have covered a large variety of qualitatively
different momentum cutoffs including exponential, algebraic, power-law, sharp
cutoffs and cutoffs with compact support. Except for $r_{{\rm opt},n}$, all
cutoffs are $C^{(\infty)}$-functions. We employ the numerical techniques
developed in \cite{Litim:2002cf,Bervillier:2007rc}.\step

%********|*********|*********|*********|*********|*********|*********|****
%********|*********| Figure 2                    |*********|*********|****
%********|*********|*********|*********|*********|*********|*********|****
\begin{center}
\begin{figure}
\unitlength0.001\hsize
\begin{picture}(1000,470)
\put(80,420){{a)}\ \ $\omega$ vs. $-\omega_0$}
\put(480,420){{d)}\ \ $\omega$ vs. $\omega_2$}
\psfig{file=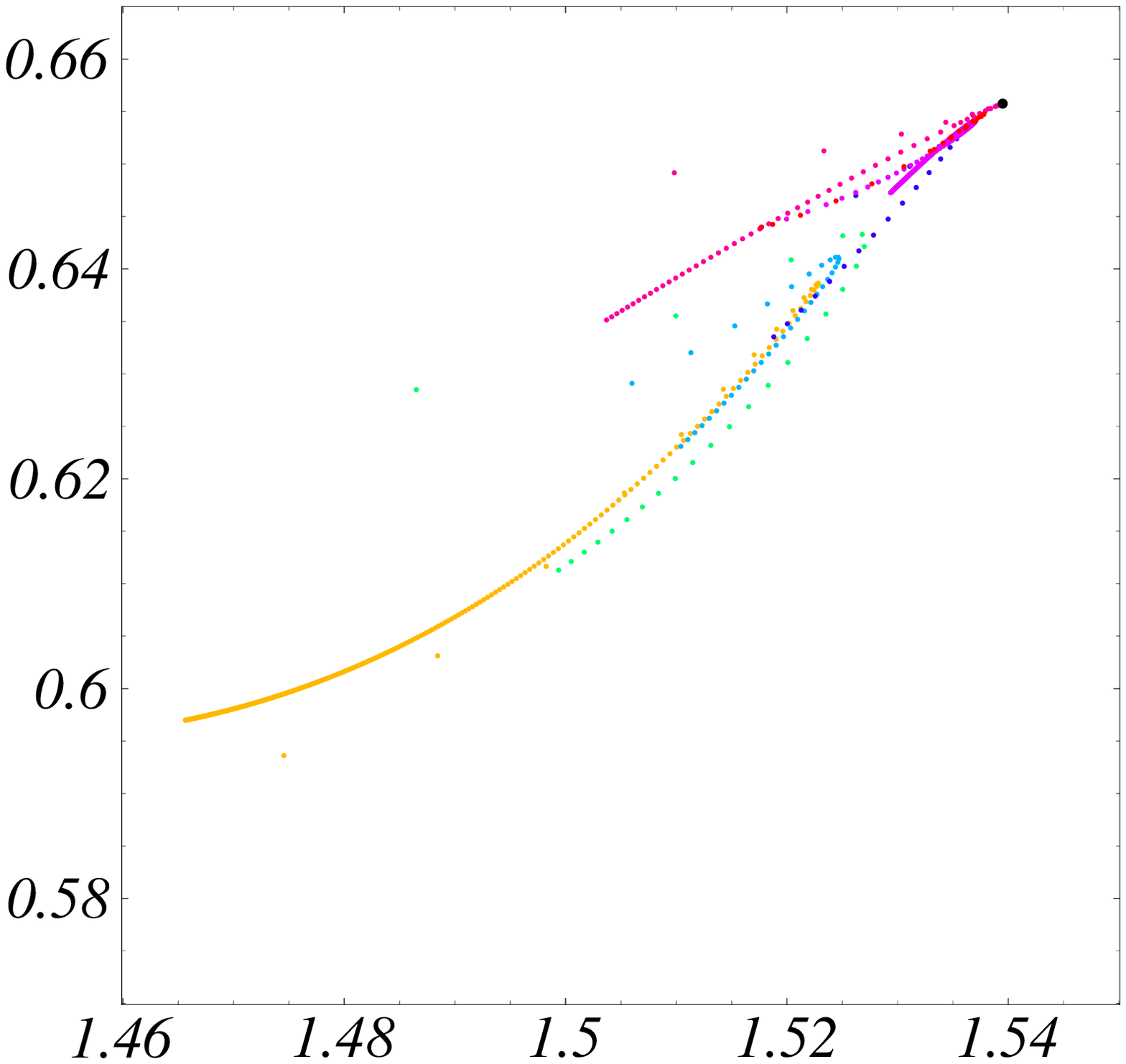,width=.4\hsize}
\psfig{file=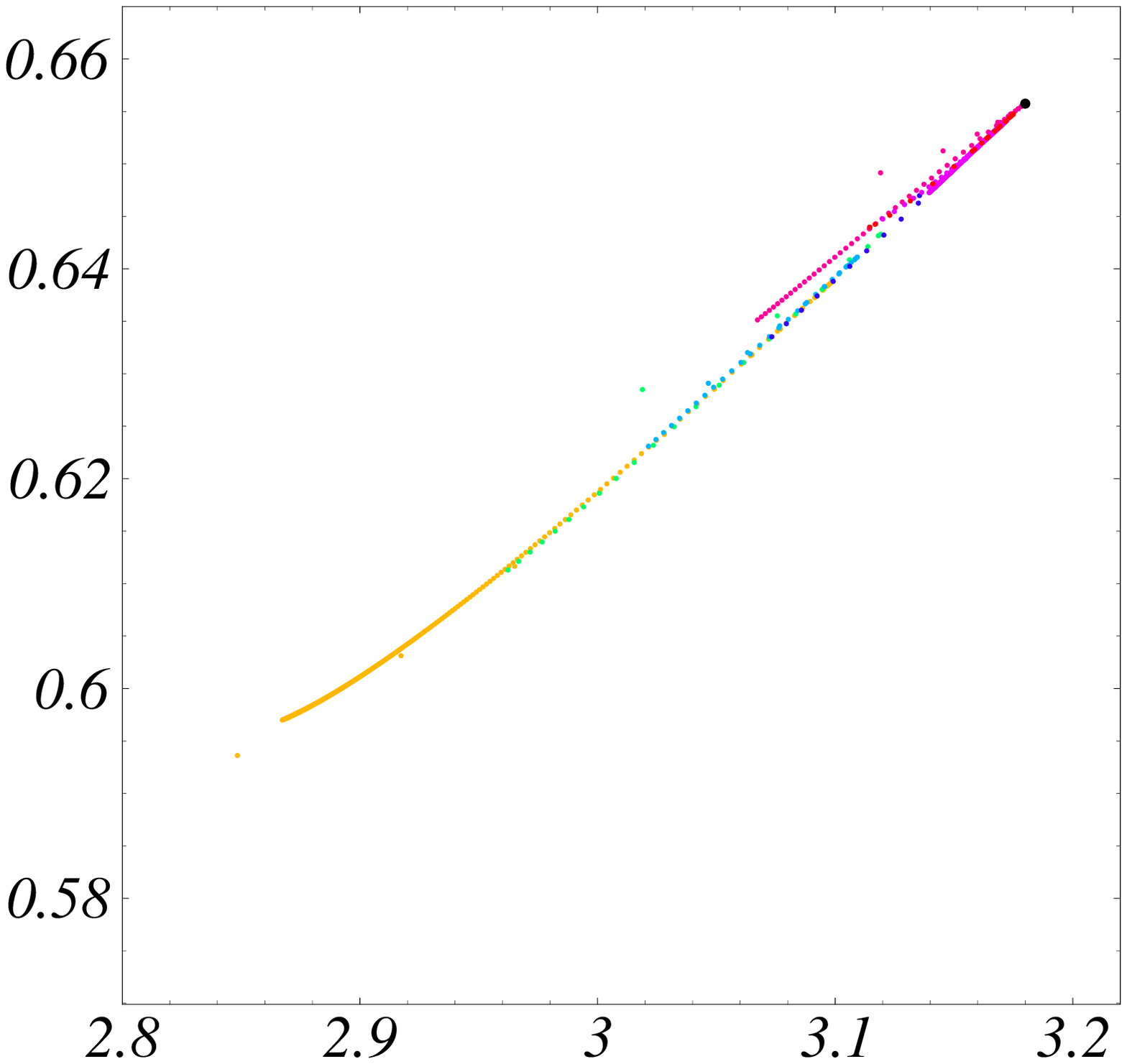,width=.4\hsize}
\end{picture}
\begin{picture}(1000,400)
\put(80,420){{b)}\ \ $\omega_2$ vs. $-\omega_0$}
\put(480,420){{e)}\ \ $\omega_2$ vs. $\omega_3$}
\psfig{file=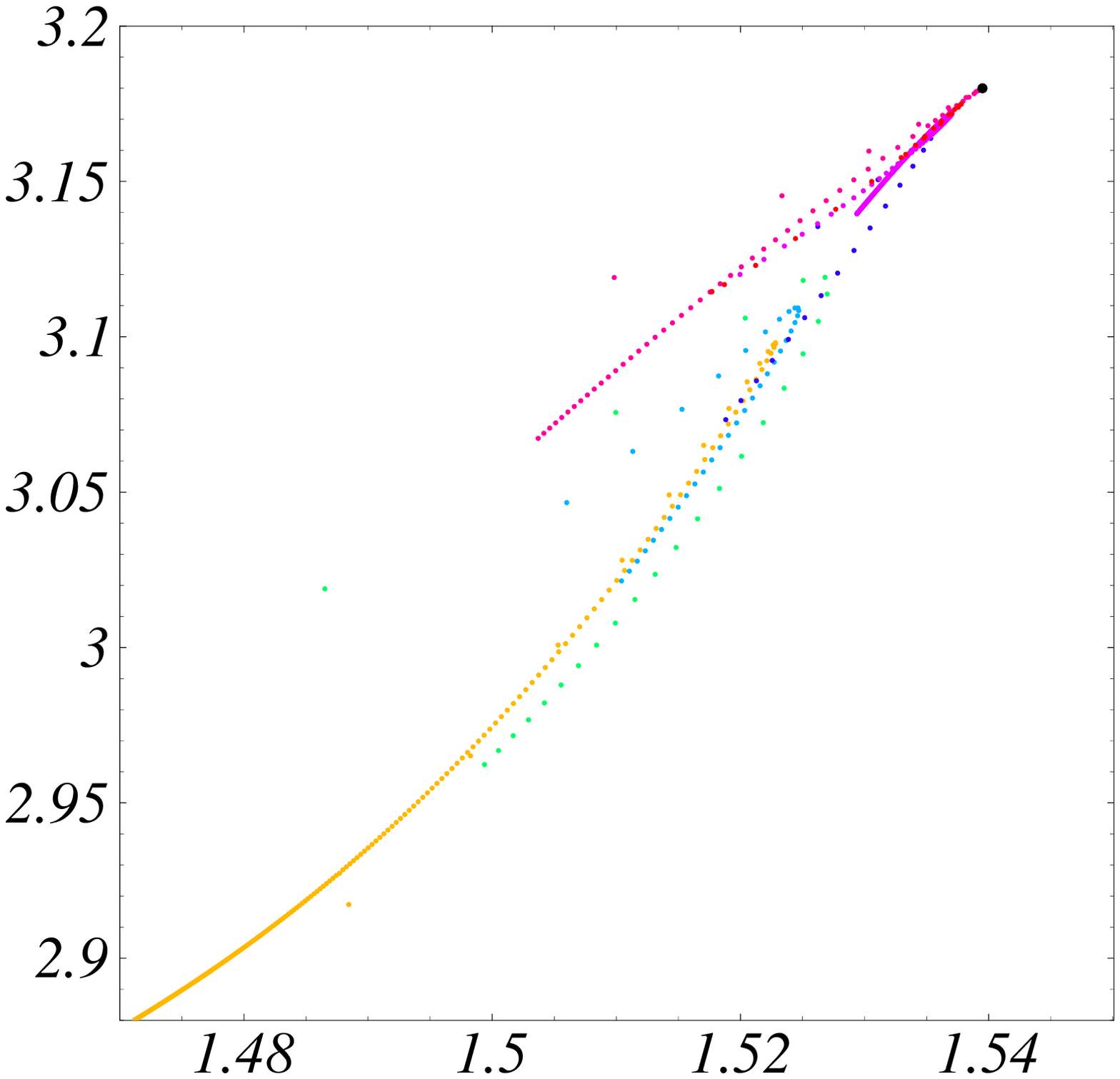,width=.4\hsize}
\psfig{file=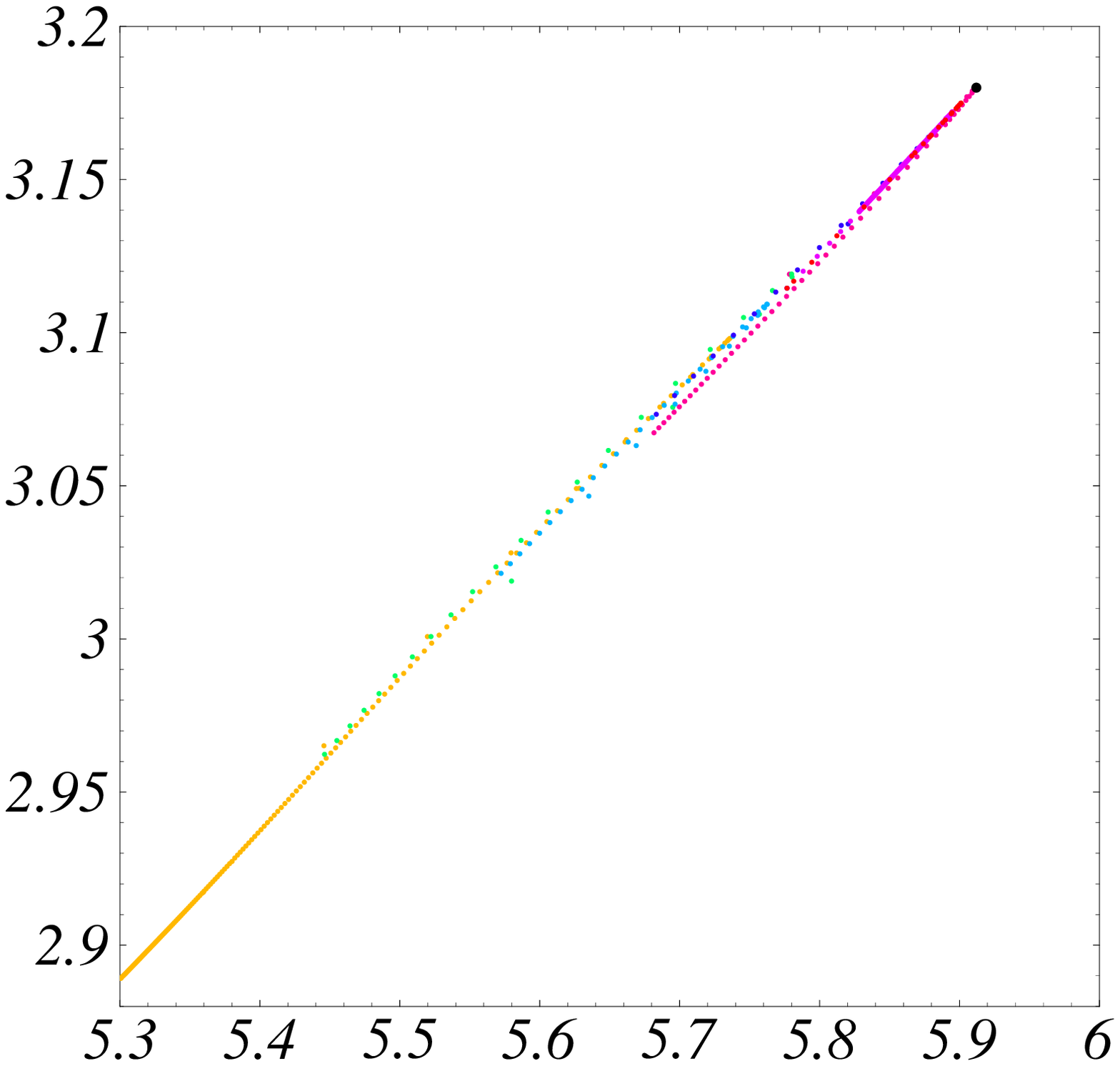,width=.4\hsize}
\end{picture}
\begin{picture}(1000,400)
\put(80,420){{c)}\ \ $-\omega_0$ vs. $\omega_3$}
\put(480,420){{f)}\ \ $\omega$ vs. $\omega_3$}
\psfig{file=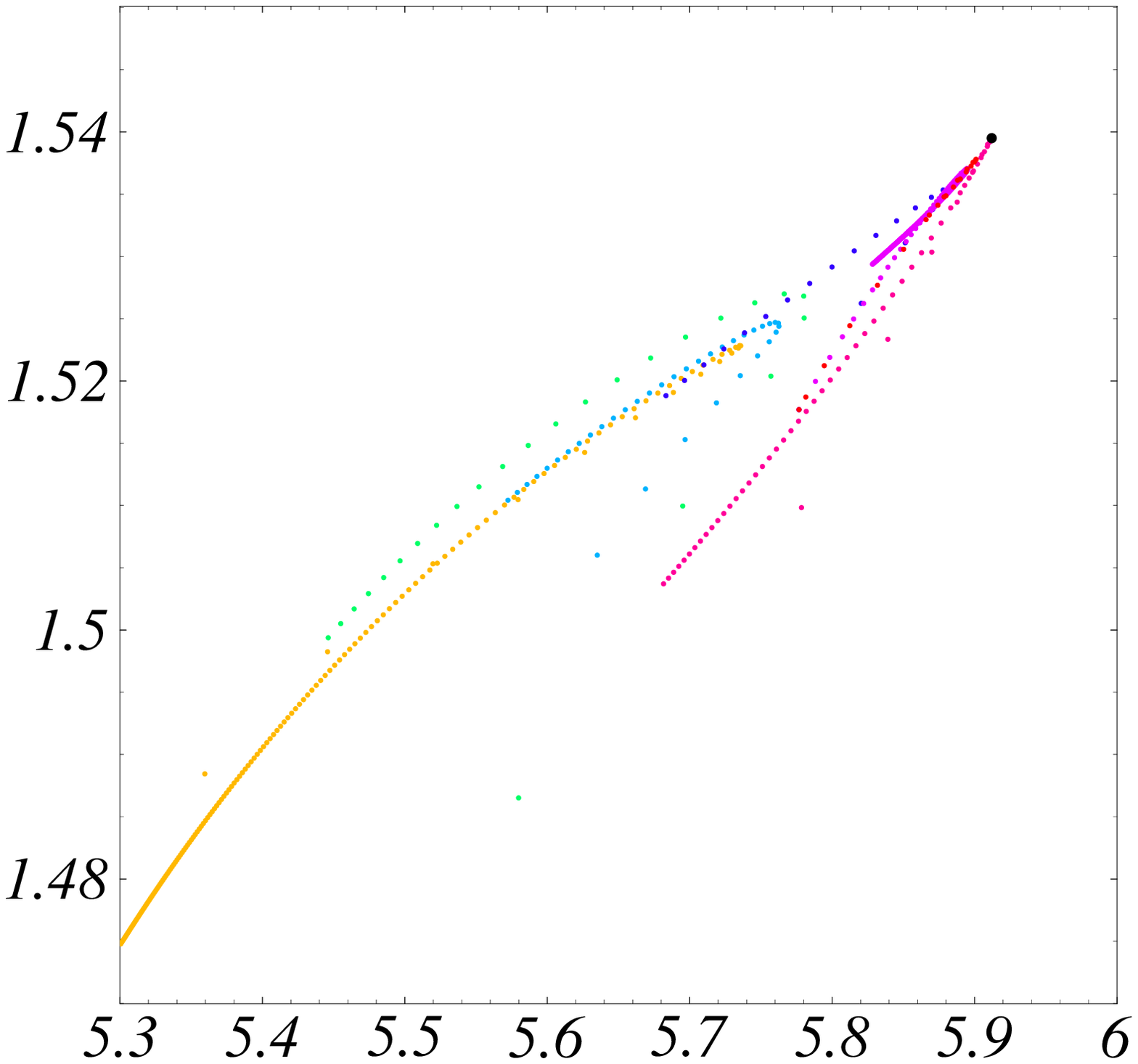,width=.4\hsize}
\psfig{file=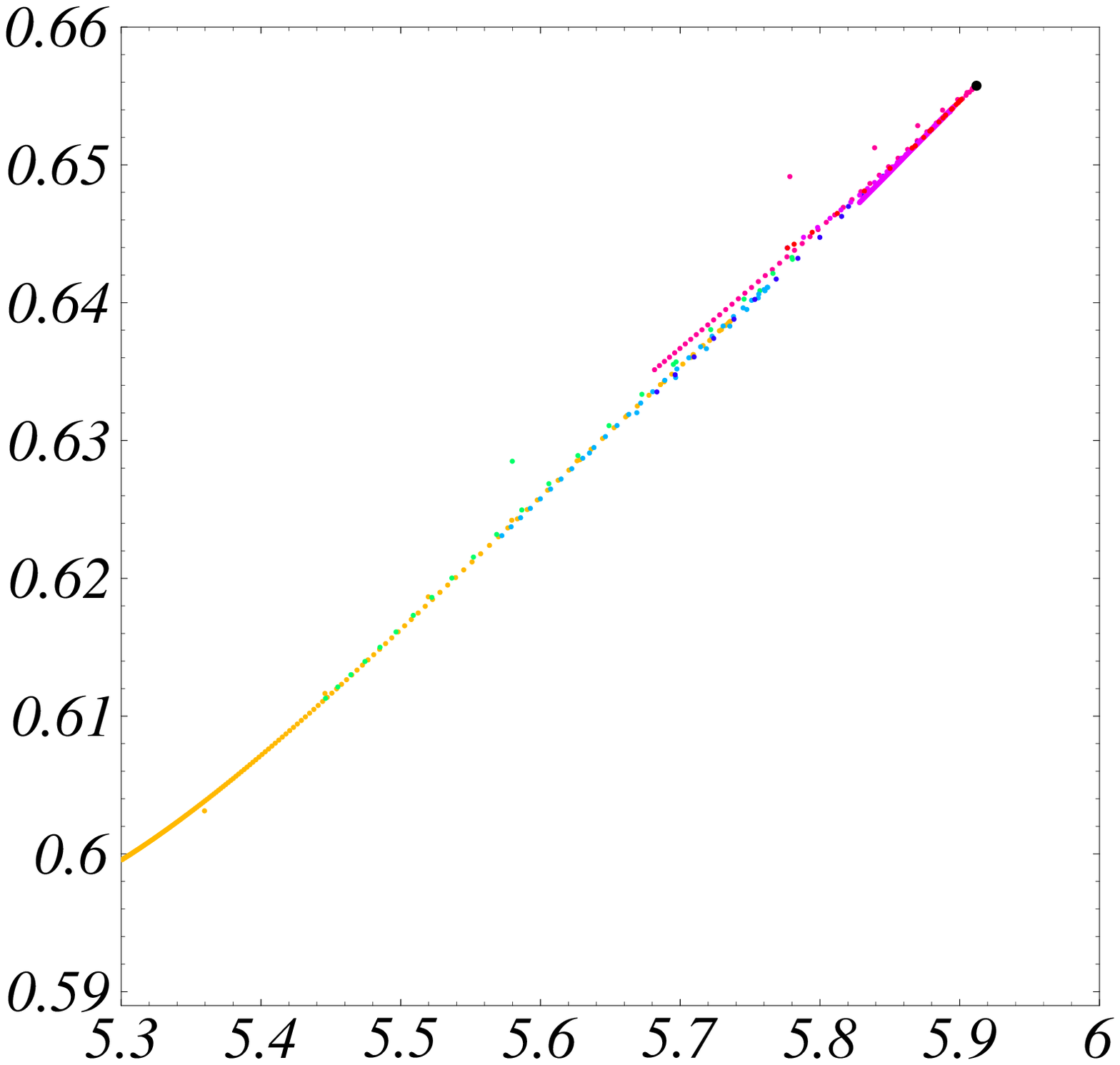,width=.4\hsize}
\end{picture}
\vskip-8.5cm
\begin{picture}(1000,470)
\put(880,1110){$r_{\rm mod}$}
\put(880,970){$r_{{\rm opt},1}$}
\put(880,820){$r_{\rm mexp}$}
\put(880,670){$(r_{{\rm PT}})$}
\put(880,530){$r_{\rm exp}$}
\put(880,385){$r_{{\rm mix}}$}
\put(880,240){$r_{\rm mix,opt}$}
\put(880,80){$r_{\rm power}$}
\hskip.84\hsize
\epsfig{file=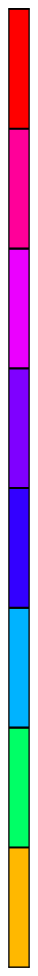,width=.0255\hsize}
\end{picture}
\vskip.5cm
 \caption{
   (colour online) Six two-dimensional projections of the four leading scaling
   exponents in the Ising universality class from the functional flow
   \eq{flowERG} for various cutoffs and coarse graining (approximately $10^3$
   data sets). Here, $\nu\equiv-1/\omega_0$. The Wilson-Polchinski result from
   the optimal flow \eq{flowOpt} (large black dot) corresponds to a local
   extremum for all scaling exponents. Data sets based on $r_{\rm power}$,
   $r_{\rm mix,opt}$, $r_{{\rm mix}}$, $r_{\rm exp}$, $r_{\rm mexp}$, $r_{\rm
     opt,1}$ and $r_{\rm mod}$; data from $r_{\rm PT}$ is included in
   Fig.~\ref{NuOmegaLog10}.
\label{fCrossCorrelation}}
\end{figure}
\end{center}
%********|*********|*********|*********|*********|*********|*********|****
\vskip-1.5cm

Our results for the universal eigenvalues at criticality are displayed in
Fig.~\ref{fCrossCorrelation} for the six two-dimensional projections of the
four-dimensional subspace $\{-\omega_0,\omega,\omega_2,\omega_3\}$ of
observables. The plot contains roughly $10^3$ data points, the different
classes of cutoffs are colour-coded.  We focus on the relevant $10\%$-vicinity
of the Wilson-Polchinski result with scaling exponents ${\cal O}_{\rm
  opt}\equiv{\cal O}(R_{\rm opt})$ from \eq{flowOpt}, indicated by a large
black dot, see \cite{Bervillier:2007rc} for the high-accuracy numerical
values. The central result of Fig.~\ref{fCrossCorrelation} is that scaling
exponents are very strongly correlated.  Despite having probed the space of
observables by many qualitatively different momentum cutoffs, we find that
only a small subset of values can actually be achieved.  The correlations
increase the closer the eigenvalues ${\cal O}$ move towards ${\cal O}_{\rm
  opt}$. In the immediate vicinity of the Wilson-Polchinski result, we only
find a very narrow ``throat'' connecting observables ${\cal O}(R)$ with ${\cal
  O}_{\rm opt}$.  For the sub-leading scaling exponents $\omega_i$, the throat
remains very narrow even further away from ${\cal O}_{\rm opt}$.  This is seen
most clearly in the correlation of $\omega_2$ with $\omega_3$ in
Fig.~\ref{fCrossCorrelation}e), as well as in the correlations of $\omega$
with both $\omega_2$ and $\omega_3$ in Fig.~\ref{fCrossCorrelation}d) and f).
In turn, for the leading exponents $\nu$, the throat opens up more rapidly
once its value is further away from $\nu_{\rm opt}$, see
Fig.~\ref{fCrossCorrelation}a), b) and c).  \bigstep

%*************************************************************************
\noindent
{\bf 7. Correlations of Eigenvalues}\\[-1ex]%\step
%*************************************************************************

The strong correlation of scaling exponents is a structural fingerprint of
Wilsonian flows \eq{flowERG}.  Since
the Wilson-Polchinski result is distinguished in the space of scaling
exponents, it is natural to normalise the data of Fig.~\ref{fCrossCorrelation}
with respect to it. We introduce the distance of any pair of scaling exponents
$(x,y)$ from the optimal result $(x_{\rm opt},y_{\rm opt})$ as
\begin{equation}\label{polar}
\rho(x,y)=\sqrt{(x_{\rm opt}-x)^2+(y_{\rm opt}-y)^2}
\equiv 10^{-N_\rho(x,y)}
\end{equation} 
We have chosen a standard metric in the space of observables (other choices
can be applied as well).  In this representation, full agreement with the
(optimal) Wilson-Polchinski result is achieved for $\rho\to0$ and
$N_\rho\to\infty$. We also introduce the angles
\begin{equation}\label{angle}
\varphi(x,y)=\arctan(x/y)\,.
\end{equation} 
The critical indices $\nu$ and $\omega_i,\,i\ge 1$, are positive numbers.
Therefore, they can cover the range $x/y\in [0,\infty]$ and
$\varphi\in[0,\s0{\pi}{2}]$, and $\rho\ge 0$ for any pair of observables
$(x,y)$.  In the subspace $(\nu,\omega)$, the extremal values $(\nu_{\rm
  opt},\omega_{\rm opt})$ have the polar coordinates $(\rho_{\rm
  opt},\varphi_{\rm opt})$, where the angle reads $\varphi_{\rm
  opt}=0.78066\cdots$ which is close to $\pi/4=0.785398\cdots$, and $\rho_{\rm
  opt}=0$. The radial distance from the origin is $\rho(0,0)=0.923002\cdots$.
\step

%*************************************************************************
% Figure 3 - polar representation
%*************************************************************************
\begin{figure}
\begin{center}
\unitlength0.001\hsize
\begin{picture}(500,500)
\put(250,455){$\uparrow$}
\put(280,450){Wilson-Polchinski}
\put(290,420){(optimal flow)}
\put(60,410){\Large $N_{\rho}(\nu,\omega)$}
\put(275,328){$\leftarrow$ hierarchical model}
\put(310,290){(Dyson, $\ell=2^{1/3}$)}
\put(100,85){sharp cutoff $\to$}
\put(40,40){Callan-Symanzik  $\to$}
\put(15,0){$0$}
\put(115,0){$\pi/8$}
\put(235,0){$\pi/4$}
\put(355,0){$3\pi/8$}
\put(470,0){${\pi}/{2}$}
\put(210,-40){\Large $\varphi(\nu,\omega)$}
\includegraphics[width=.5\hsize]{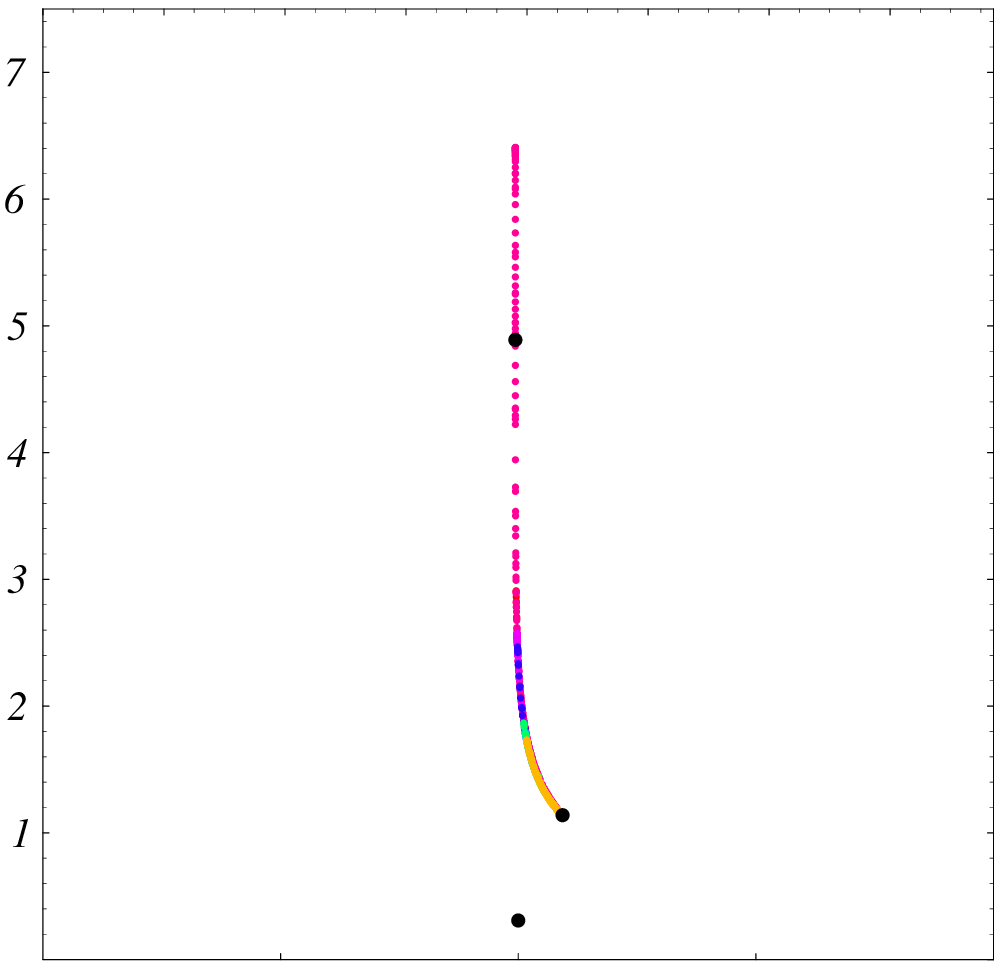}
\end{picture}
\vskip-7.5cm
\unitlength0.001\hsize
\begin{picture}(1000,470)
\put(810,460){$r_{\rm mod}$}
\put(810,395){$r_{{\rm opt},1}$}
\put(810,335){$r_{\rm mexp}$}
\put(810,270){$(r_{{\rm PT}})$}
\put(810,210){$r_{\rm exp}$}
\put(810,150){$r_{{\rm mix}}$}
\put(810,90){$r_{\rm mix,opt}$}
\put(810,30){$r_{\rm power}$}
\hskip.78\hsize
\epsfig{file=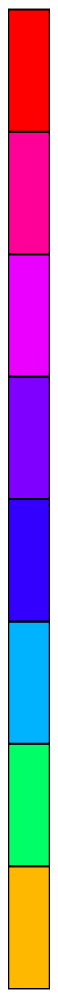,width=.021\hsize}
\end{picture}
\vskip1cm
\caption{ 
  (colour online) Distance $N_\rho$ of the pair of scaling exponents
  $(\nu,\omega)$ from the optimal Wilson-Polchinski values $(\nu_{\rm
    opt},\omega_{\rm opt})$ in the representation \eq{polar}, \eq{angle}.
  Only a narrow range of angles $\varphi(\nu,\omega)$ in the vicinity of
  $\varphi\approx \pi/4$ is achieved by the data. Many data points are nearly
  degenerate. Data from functional flows in the local potential approximation
  \eq{flowERG}; same data sets and colour coding as in
  Fig.~\ref{fCrossCorrelation}, plus further high resolution data points from
  $r_{{\rm opt},1}$ in the close vicinity of $(\nu_{\rm opt},\omega_{\rm
    opt})$.  Results from the sharp cutoff limit, the Callan-Symanzik type
  flow (with $R_k\sim k^2$) and Dyson's hierarchical model (with
  $\ell=2^{1/3}$) are also indicated (black dots). The Wilson-Polchinski
  (optimal flow) result corresponds to $\varphi=\varphi_{\rm opt}$ and
  $N_\rho\to\infty$.
  \label{pNuOmegaDistance} }\vskip-.5cm
\end{center}
\end{figure}
%*************************************************************************

%*************************************************************************
% Figure 4 
%*************************************************************************
\begin{figure}
\begin{center}
\unitlength0.001\hsize
\begin{picture}(500,500)
\put(340,390){{\Large $N_{\rho}(\nu,\omega)$}}
\put(37,460){$\uparrow$}
\put(70,470){Wilson-Polchinski}
\put(75,440){(optimal flow)}
\put(70,328){$\leftarrow$ hierarchical model}
\put(105,300){(Dyson, $\ell=2^{1/3}$)}
\put(345,115){sharp cutoff}
\put(405,90){$\leftarrow$}
\put(80,40){$\leftarrow$ Callan-Symanzik}
\put(235,-40){\Large $\nu/\omega$}
\includegraphics[width=.5\hsize]{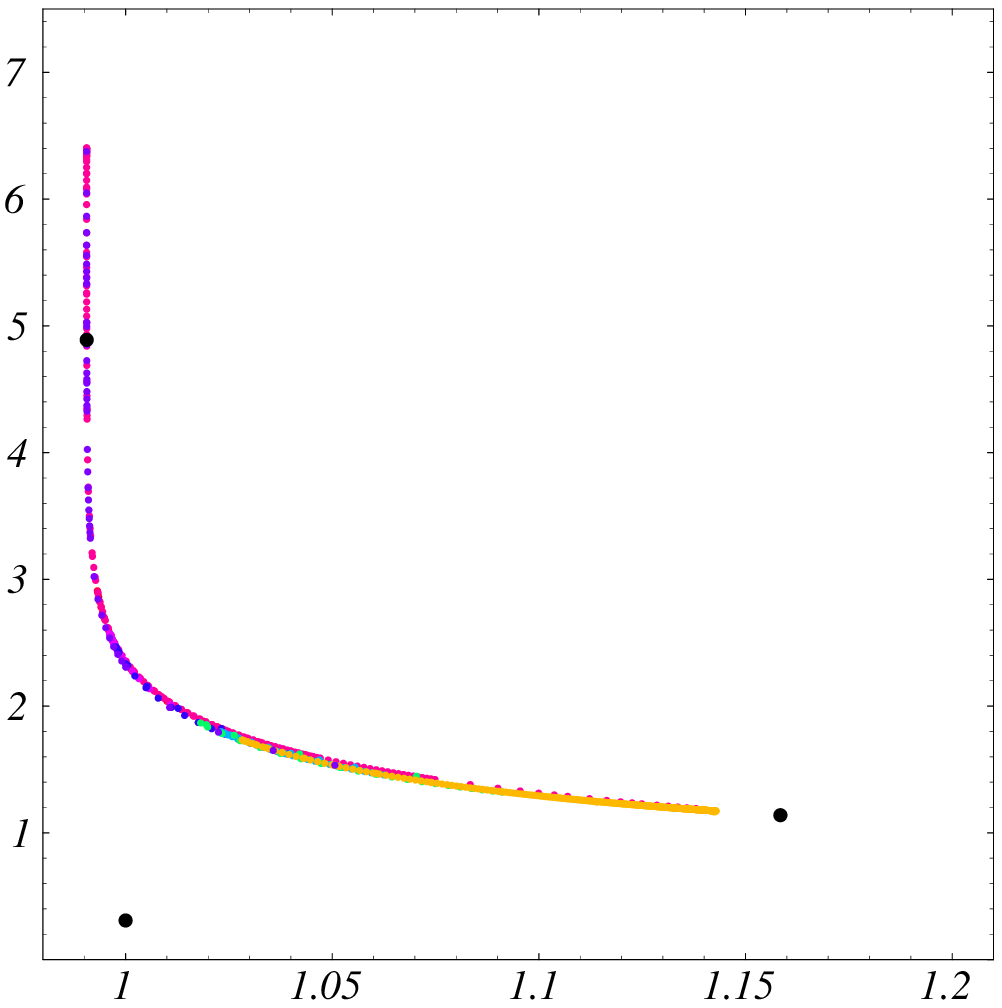}
\end{picture}
\vskip-7.5cm
\unitlength0.001\hsize
\begin{picture}(1000,470)
\put(810,460){$r_{\rm mod}$}
\put(810,395){$r_{{\rm opt},1}$}
\put(810,335){$r_{\rm mexp}$}
\put(810,270){$r_{{\rm PT}}$}
\put(810,210){$r_{\rm exp}$}
\put(810,150){$r_{{\rm mix}}$}
\put(810,90){$r_{\rm mix,opt}$}
\put(810,30){$r_{\rm power}$}
\hskip.78\hsize
\epsfig{file=pLegendeL.eps,width=.021\hsize}
\end{picture}
\vskip1.5cm
\caption{\label{NuOmegaLog10} 
  (colour online) Magnification of Fig.~\ref{pNuOmegaDistance} in the vicinity
  of $\varphi\approx \pi/4$ where $\nu/\omega\approx 1$. The data points for
  the distance $N_\rho(\nu,\omega)$ as a function of $\nu/\omega$ remain
  highly degenerate.  Same data sets and colour coding as in
  Fig.~\ref{pNuOmegaDistance}, plus additional high resolution data points
  from background field flows \eq{flowPT} using $r_{{\rm PT},m}$ with
  $m<\s052$.  Results from the sharp cutoff limit, the Callan-Symanzik type
  flow (with $R_k\sim k^2$) and Dyson's hierarchical model (with
  $\ell=2^{1/3}$) are also indicated (black dots). The Wilson-Polchinski
  (optimal flow) result corresponds to $\nu_{\rm opt}/\omega_{\rm
    opt}=0.9905692\cdots$ and $N_\rho\to\infty$.}\vskip-.5cm
\end{center}
\end{figure}
%*************************************************************************

In the representation \eq{polar}, we can study the close vicinity of the
Wilson-Polchinski result.  In Fig.~\ref{pNuOmegaDistance}, we display our data
points as functions of the angles $\varphi$, and their distance from
$\rho_{\rm opt}$ in a semi-logarithmic basis. It is noteworthy that only a
very narrow range of angles $\varphi$ is actually achieved by the data,
despite the fact that large fractions of the underlying space of momentum
cutoffs is covered.  Also, and in contrast to Fig.~\ref{fCrossCorrelation}a),
many data points are degenerate in the representation $(N_\rho,\varphi)$. A
priori, the Wilson-Polchinski value $(\rho_{\rm opt},\varphi_{\rm opt})$ could
have been approached along many different paths. Instead, we find that only a
narrow range of $(N_\rho,\varphi)$-values is achieved for arbitrary
momentum cutoff.\step  

This pattern is further highlighted in Fig.~\ref{NuOmegaLog10}, where we have
magnified the non-trivial range of data sets from functional flows
\eq{flowERG}. In addition, we have added data points from the background field
flow \eq{flowPT} using the cutoff $r_{{\rm PT},m}$ for $m<\s052$. It is
remarkable that these data sets display the same pattern as the data from
\eq{flowERG}. Our results from this and the preceeding section are summarised
as follows:\step

{\it Extremum.---} The Wilson-Polchinski (optimal flow) result in
Fig.~\ref{fCrossCorrelation} corresponds to an extremum in the space of
physical observables with ${|\omega_i|} \le |\omega_{i,{\rm opt}}|$ for all
obervables in the vicinity of ${\cal O}_{\rm opt}$.  The extremum is local,
because the exponents approach $\omega_i=2i-1$, $i\ge 0$, for very soft
(Callan-Symanzik-type) momentum cutoffs \cite{Litim:2002cf}. For the
eigenvalue products $\Pi_{i=0}^n(\omega_i/\omega_{i,{\rm opt}})$, the
Wilson-Polchinski extremum is a global one.\step

{\it Uniqueness.---} Our result indicates that the correlations of
eigenvalues at the Wilson-Pochinski result are strongest, in the sense that
any flow of the form \eq{flowERG} with the exponent $\nu(R)=\nu_{\rm opt}$
automatically also agrees with the Wilson-Polchinski result in all other
observables ${\cal O}(R)={\cal O}_{\rm opt}$. In general, for $\nu(R)>\nu_{\rm
  opt}$, this is clearly not the case.\step

{\it Redundancy.---} 
The eigenvalue correlations are so strong that the first two scaling exponents
$\nu(R)$ and $\omega(R)$, for a given $R$, contain enough information to fix
the remaining observables on the percent level or below.  These `dynamical'
constraints point at a major redundancy of \eq{flowERG} with respect to the
underlying momentum cutoffs $R$. A relevant parameter has been identified
previously. The gap $\min_{y\ge 0} y(1+r)$ for normalised cutoffs
\cite{Litim:2000ci}, when maximised, leads towards the Wilson-Polchinski
result \cite{Litim:2001up,Litim:2001fd,Litim:2002cf}.\step

{\it Optimised observables.---} Previous reasonings in favour of an
optimisation only invoked properties of the underlying flow \eq{ERG},
$e.g.$~its convergence, locality, stability and boundedness, allowing for
improved physical predictions.  This has been exemplified quantitatively for
the observable $\nu(R)$ which obeys $1\ge\nu(R)\ge\nu_{\rm opt}$
\cite{Litim:2002cf}, where the lower bound $\nu_{\rm opt}$ is closest to the
physical result \cite{Litim:2005us}.  Fig.~\ref{fCrossCorrelation} now shows
that this pattern extends to subleading eigenvalues. This equally extends to
asymmetric corrections-to-scaling \cite{Litim:2003kf}.  Therefore, one may
turn the original reasoning around and argue that -- because of the extremum
property of the observables ${\cal O}(R)$ -- an extremisation of the
functional flow along the lines discussed in
\cite{Litim:2000ci,Litim:2001up,Pawlowski:2005xe}, or similar, should
naturally lead towards the values ${\cal O}_{\rm opt}$. Stated differently,
Figs.~\ref{fCrossCorrelation}-\ref{NuOmegaLog10} show that observables derived
from \eq{ERG} admit an optimisation.\step

Finally, we note that the data point from Dyson's hierarchical model with
$\ell=2^{1/3}$ -- as plotted in Figs.~\ref{pNuOmegaDistance}
and~\ref{NuOmegaLog10} -- nicely fits into the set of data points covered by
functional flows, extending the link observed in Sec.~5 beyond the leading
exponent. This observation is addressed quantitatively in the following
section.\bigstep

%*************************************************************************
\noindent
{\bf 8. Matching Beyond the Leading Exponent}\\[-1ex]%\step
%*************************************************************************

To further substantiate our conjecture that hierarchical models could be
mapped onto functional flows, we have to show quantitatively that results from
hierarchical model are reproduced by specific functional flows.  Here, we
study the close vicinity of the Wilson-Polchinski result ${\cal O}_{\rm opt}$,
where the correlations are strongest, see Figs.~\ref{fCrossCorrelation}
and~\ref{pNuOmegaDistance}.  We have to restrict our search to Dyson's
hierarchical model, where high-accuracy data for the first sub-leading scaling
exponent $\omega$ is available. No subleading exponents have been computed for
Wilson's model.\step

For the numerical analysis, we introduce additional classes of momentum
cutoffs $R$ which contain the optimal flow \eq{flowOpt} in some limit. In
addition to the two-parameter family of cutoffs $r_{{\rm opt},n}$, we also
study the cutoff $r_{\rm compact}=y^{-1}\exp[-e^{-1/y}/(b-y) ]\theta(b-y)$ for
$b>0$ which is $C^{(\infty)}$, and the cutoff $r_{\rm
  int}=\exp(-y)\theta(1-y)\theta(y-b)$ with $b\in [0,1]$, which is effective
for a finite interval of momenta $q^2\in [bk^2,k^2]$. In the limit $b\to 0$
$(b\to 1)$, the corresponding flows are equivalent to \eq{flowOpt}.  Hence,
$r_{{\rm opt},n}$, $r_{\rm compact}$ and $r_{\rm int}$ parametrise
substantially different classes of cutoffs. More generally, there are
infinitely many cutoffs $R_k$ leading to scaling exponents identical with
${\cal O}_{\rm opt}$, and the examples provided above serve to
illustrate this.\step

%********|*********|*********|*********|*********|*********|*********|****
%********| Table 1
{\small
\begin{table}
\begin{center}
\begin{tabular}{lllll}
\\[-2ex]
\multicolumn{1}{c}{method}
&\multicolumn{1}{c}{cutoff}
&\multicolumn{1}{c}{parameter}
&\multicolumn{1}{c}{$\nu$}
&\multicolumn{1}{c}{$\omega$}\\[.5ex]\hline\\[-2ex]
${}\quad$ hierarchical model ${}\quad$
&\multicolumn{1}{c}{Dyson}
&$(\ell=1)$%\multicolumn{2}{c}{$(\ell=2^{1/3})$}
&\quad 0.649\,561\,773\,880${}^a$
\quad 
&\quad 0.655\,745\,939\,193${}^a$
\quad \\[.5ex]\hline\\[-2ex]
& $r_{{\rm opt},n}$
&($n=1,\,b=1$)
&\quad 0.649\,561\,773\,880${}^a$&\quad 0.655\,745\,939\,193${}^a$
\\[.5ex]
\multicolumn{1}{c}{functional RG}
& $r_{\rm compact}$
&($b\to 0$)
&\quad 0.649\,561\,773\,880
&\quad 0.655\,745\,939\,193
\\[.5ex]
& $r_{\rm int}$
&($b\to 1$)
&\quad 0.649\,561\,773\,880
&\quad 0.655\,745\,939\,193
\\[.5ex]
&$r_{{\rm PT},m}$
&($m=5/2$)
&\quad 0.649\,561\,773\,880${}^a$
&\quad 0.655\,745\,939\,193${}^a$
\\[1ex]
\\[-2ex]
\end{tabular}
\caption{Matching scaling exponents $\nu$ and $\omega$ from continuous
  hierarchical transformations with functional flows. Results agree at least
  to the order $10^{-12}$. Data from this work, and from $a)$
  \cite{Bervillier:2007rc}.}
\vskip.5cm
\label{Comparison1}
\end{center}
\end{table}
}
%********|*********|*********|*********|*********|*********|*********|****

At $\ell=1$, Dyson's hierarchical transformation is continuous, and the
scaling exponents are equivalent to those from the optimal flow \eq{flowOpt}
and the Wilson-Polchinski flow \eq{flowWP}. In Tab.~\ref{Comparison1}, we
compare exponents from different functional flows. We confirm numerically, and
with high accuracy, that the cutoffs $r_{{\rm opt},n}$, $r_{\rm compact}$ and
$r_{\rm int}$ lead to the Wilson-Polchinski result for specific parameter
values.\step

At $\ell=2^{1/3}$, Dyson's hierarchical transformation is discrete. The
reference data reads $\nu_{{}_{\rm DHM}}= 0.649570$ and $\omega_{{}_{\rm
    DHM}}= 0.655736$ \cite{Godina:1998uz}.\footnote{In \cite{Godina:1998uz},
  high-accuracy results at $\ell=2^{1/3}$ have been given for
  $\gamma=2\,\nu$ %gamma=1.2991407301586
  and $\Delta=\nu\,\omega$ %Delta=0.4259468589881
  (and $\eta=0)$ with 13 significant digits. They imply $\nu_{\rm DHM}
  =0.649570365\cdots$ and $\omega_{\rm DHM}=0.655736286\cdots$. For the
  present study, only the first six figures are required.} These values differ
only at the order $10^{-5}$ from the optimal (Wilson-Polchinski) result, and
are therefore sufficiently close to ${\cal O}_{\rm opt}$ to confirm or refute
the correlations observed in the previous section.
Figs.~\ref{pNuOmegaDistance} and \ref{NuOmegaLog10} indicate that the result
from Dyson's hierarchical model is fully matched by functional flows.  Our
numerical results are given in Tab.~\ref{Comparison2}; brackets indicate that
a digit is possibly affected by numerical errors.  We have found several sets
of parameter values, such that the scaling exponents agree with Dyson's model
to order $10^{-6}$.  More importantly, the momentum cutoffs are quite
different.  Hence, our analysis also confirms the strong correlation of
scaling exponents in the immediate vicinity of the Wilson-Polchinski result.
Based on the eigenvalue correlations within functional flows, we conjecture
that the subleading eigenvalues $\omega_i$ with $i\ge 2$ of Dyson's model at
$\ell=2^{1/3}$ also agree to the corresponding accuracy with the values
implied through the functional flows in Tab.~\ref{Comparison2}. \step

Our results based on the proper-time flow \eq{flowPT} with $r_{\rm PT}$ has
also been given in Tab.~\ref{Comparison2}. Full agreement is achieved either
with the exponent $\nu_{{}_{\rm DHM}}$ or the subleading exponent
$\omega_{{}_{\rm DHM}}$, but not with both of them. Once one of them is
matched, the deviation in the other observable is of the order $10^{-5}$. The
relevant parameters are $m<\s052$, the regime where \eq{flowPT} is mapped onto
\eq{flowERG} \cite{Litim:2002hj}.  Therefore, the values in
Tab.~\ref{Comparison2} reflects well the range covered by standard Wilsonian
flows \eq{flowERG}.  We expect that full agreement is achieved for proper-time
flows which are linear combinations of \eq{flowPT} for different $m$, but we
did not attempt to do so here.\step

%********|*********|*********|*********|*********|*********|*********|****
%********| Table 2
%********|*********|*********|*********|*********|*********|*********|****
\begin{table}
\begin{center}
\begin{tabular}{lllll}
\\[-2ex]
\multicolumn{1}{c}{method}
&\multicolumn{1}{c}{cutoff}
&\multicolumn{1}{c}{parameter}
&\multicolumn{1}{c}{$\nu$}
&\multicolumn{1}{c}{$\omega$}\\[.5ex]\hline\\[-2ex]
${}\quad$ hierarchical model ${}\quad$
&\multicolumn{1}{c}{Dyson}
&$(\ell=2^{1/3})$%\multicolumn{2}{c}{$(\ell=2^{1/3})$}
&\quad 0.649\,570${}^b$\quad 
&\quad 0.655\,736${}^b$\quad \\[.5ex]\hline\\[-2ex]
& $r_{{\rm opt},n}$
&($n=1,\,b=1.048$)
&\quad 0.649\,570(9) 
&\quad  0.655\,736(6)\\[.5ex]
& $r_{{\rm opt},n}$
&($n=1,\,b=0.9545$)
&\quad 0.649\,570(9)
&\quad 0.655\,736(9)\\[.5ex]
& $r_{{\rm opt},n}$
&($n=1.135,\, b=1$)
&\quad 0.649\,570(6)
&\quad 0.655\,736(8)\\[.5ex]
& $r_{{\rm opt},n}$
&($n=1.1,\, b=1.028$)
&\quad 0.649\,570(6)
&\quad 0.655\,736(8)\\[.5ex]
\multicolumn{1}{c}{functional RG}
& $r_{\rm compact}$
&($b=0.04775$)
&\quad 0.649\,570(9)
&\quad  0.655\,736(9)\\[.5ex]
& $r_{\rm int}$
&($b= 0.944$)
&\quad 0.649\,570(9)
&\quad 0.655\,736(8)
\\[.5ex]
& $r_{\rm int}$
&($b= 0.9444$)
&\quad 0.649\,570(7)
&\quad 0.655\,736(9)
\\[.5ex]
&$r_{{\rm PT},m}$
&($m=2.499785$)
&\quad 0.649\,564(9)
&\quad 0.655\,736(1)
\\[.5ex]
&$r_{{\rm PT},m}$
&($m=2.49944$)
&\quad 0.649\,570(1)
&\quad 0.655\,720(6)
\\[1ex]
\end{tabular}
\caption{Matching scaling exponents $\nu$ and $\omega$ from discrete
  hierarchical transformations with functional flows. Results agree to the
  order $10^{-6}$ for all cutoffs except the proper-time flow, which matches
  up to the order $10^{-5}$. Data from this work, and from $b)$
  \cite{Godina:1998uz}.}
\label{Comparison2}
\end{center}
\end{table}
%********|*********|*********|*********|*********|*********|*********|****

In summary, we have provided numerical maps from several functional flows onto
Dyson's model at a non-trivial $\ell\neq 1$, with an accuracy of the order
$10^{-6}$. The set of achievable values for scaling exponents from functional
flows in the close vicinity of the optimal result is just wide enough to
accommodate for the data from Dyson's model. This is a non-trivial result,
also showing that the $\ell$-dependence of Dyson's model and the
$R_k$-dependence of functional flows are very intimately related.  Based on
our results for $\nu$ and $\omega$ at $\ell\neq 1$, and on continuity in
$\ell$, we expect that this map extends to other universal quantities in the
same approximation, analogous to the full map which is known for $\ell=1$.
Data for further symmetric and asymmetric corrections-to-scaling exponents,
once available, will allow for additional checks of this picture. Full
equivalence is guaranteed as soon as an explicit link in the form
$\ell=\ell(R_k)$ or $R_k(q^2)=R_k(q^2,\ell)$ is furnished. For the local
potential approximation, our results indicate that this map, if it exists, is
not unique.\bigstep

%********|*********|*********|*********|*********|*********|*********|****
\noindent
{\bf 9. Discussion and Conclusions}\\[-1ex]%\step
%********|*********|*********|*********|*********|*********|*********|****

Establishing equivalences between implementations of Wilson's renormalisation
group as different as discrete hierarchical models of lattice scalar fields on
one side and continuous functional flows on the other, allows for new views
and insights on the respective formalisms and on the underlying physics.
Previously, equivalences were known only in the limit where the hierarchical
transformation becomes continuous.  In this paper, based on similarities in
the dependences related to the underlying coarse-graining, we have extended
this link towards discrete hierarchical transformations. This correspondence
shows that continuous RG flows \eq{ERG} are sensitive to implicit
discretisation effects via the momentum cutoff. \step

Specifically, for the $3d$ Ising universality class, we have compared the
formalisms on the level of scaling exponents. Their dependence on the
step-size parameter $\ell$ within Dyson's hierarchical model \eq{DHM} is
qualitatively and quantitatively similar to their dependence on the momentum
cutoff $R$ within functional flows \eq{flowERG}.  In either case, scaling
exponents are bounded by the Wilson-Polchinski values obtained for $\ell \to
1$ and $R\to R_{\rm opt}$.  Once the hierarchical transformations are
discrete, $\ell\neq 1$, slight variations in all known scaling exponents from
Dyson's model are matched by functional flows with non-optimal momentum
cutoffs $R\neq R_{\rm opt}$. This is quite remarkable, particularily in view
of the strong eigenvalue correlations found amongst functional flows. In this
light, the optimisation of functional flows with $R\to R_{\rm opt}$ can now,
alternatively, be viewed as the removal of discretisation effects, at least to
leading order in a derivative expansion as studied here. It will be
interesting to contrast these findings with the construction of improved or
perfect actions on the lattice.  More generally, it is conceivable that
Dyson's model for arbitrary $\ell$ is mapped by functional flows on a
fundamental level beyond the numerical map provided for $\ell=2^{1/3}$.  An
explicit map would be very welcome, also in view of linking hierarchical
models to a path integral representation of the theory.  In Wilson's
hierarchical model \eq{WHM}, the range covered by the leading scaling exponent
indicates that a partial map onto functional flows exists, though only for a
restricted domain of $\ell$-values. Interestingly, the overlap with background
field flows is even larger. Whether these maps extend beyond the leading
exponent cannot be settled presently due to a lack of data for subleading
exponents from Wilson's model.  \step

In addition, we found a distinct correlation of scaling exponents from
functional flows \eq{flowERG}.  The eigenvalue spectrum, a fingerprint of the
physics in a local potential approximation, is severely constrained and
achieves the Wilson-Polchinski values as an extremum.  Furthermore, the full
space of physical observables is described by very few parameters only,
instead of the infinitely many moments of the momentum cutoff.  Resolving this
redundancy should prove useful for studies of $e.g.$~non-trivial momentum
structures and higher orders in the derivative expansion.  Finally this
pattern amongst physical observables highlights the extremum property of
cutoffs leading to the optimal Wilson-Polchinski result
\cite{Litim:2000ci,Litim:2001up,Litim:2001fd,Pawlowski:2005xe}.  We expect
that the intimate link between optimised flows on one side, and extremum
points in the space of observables on the other, persists in more complex
theories. This observation will prove useful for studies in QCD and quantum
gravity, where an appropriate optimisation is even more important to extract
the relevant physics.  \step

{\it Acknowledgements.---} This work is supported by an EPSRC Advanced
Fellowship.  I thank Y.~Meurice for e-mail correspondence, the Galileo
Galilei Institute for Theoretical Physics for hospitality, and the INFN for
partial support.


\begin{thebibliography}{99}


%\cite{Zinn-Justin:1989mi}
\bibitem{Zinn-Justin:1989mi}
J.~Zinn-Justin,
{\it Quantum Field Theory And Critical Phenomena},
{Oxford, Clarendon (1989)}.%

%\cite{Wilson:1973jj}
\bibitem{Wilson:1973jj}
  K.~G.~Wilson and J.~B.~Kogut,
  %``The Renormalization group and the epsilon expansion,''
  Phys.\ Rept.\  {\bf 12} (1974) 75.
  %%CITATION = PRPLC,12,75;%%

%\cite{Tetradis:1995br}
\bibitem{Tetradis:1995br}
  N.~Tetradis and D.~F.~Litim,
  %``Analytical Solutions of Exact Renormalization Group Equations,''
  Nucl.\ Phys.\ B {\bf 464} (1996) 492
  [hep-th/9512073],
  %%CITATION = HEP-TH 9512073;%%

%\cite{Morris:1998da}
\bibitem{Morris:1998da}
  T.~R.~Morris,
  %%``Elements of the continuous renormalization group,''
  Prog.\ Theor.\ Phys.\ Suppl.\  {\bf 131} (1998) 395
  [hep-th/9802039].%;\\
  %%CITATION = HEP-TH 9802039;%%

%\cite{Litim:1998yn}
\bibitem{Litim:1998yn} 
D.~F.~Litim, {\it Wilsonian flow equation and thermal
    field theory}, in: U.~Heinz (Ed.), {\it Thermal
    Field Theories and their Applications}, [hep-ph/9811272].%;\\
  %%CITATION = HEP-PH 9811272;%%

%\cite{Litim:1998nf}
\bibitem{Litim:1998nf} 
D.~F.~Litim and J.~M.~Pawlowski, in {\it The Exact
  Renormalization Group}, Eds.~Krasnitz et al, World Sci (1999) 168 
   %``On gauge invariant Wilsonian flows,''
  %
  [hep-th/9901063].%;\\
  %%CITATION = HEP-TH 9901063;%%

%\cite{Bagnuls:2000ae}
\bibitem{Bagnuls:2000ae}
  C.~Bagnuls and C.~Bervillier,
  %``Exact renormalization group equations: An introductory review,''
  Phys.\ Rept.\  {\bf 348} (2001) 91
  [hep-th/0002034].%;\\
  %%CITATION = HEP-TH 0002034;%%

%\cite{Berges:2000ew}
\bibitem{Berges:2000ew}
  J.~Berges, N.~Tetradis and C.~Wetterich,
   %``Non-perturbative renormalization flow in quantum field theory and
   %statistical physics,''
  %
  Phys.\ Rept.\  {\bf 363} (2002) 223 [hep-ph/0005122].%;\\
  %%CITATION = HEP-PH 0005122;%%

%\cite{Polonyi:2001se}
\bibitem{Polonyi:2001se}
  J.~Polonyi,
   %``Lectures on the functional renormalization group method,''
  %
  Central Eur.\ J.\ Phys.\  {\bf 1} (2004) 1
  [hep-th/0110026].%;\\
  %%CITATION = HEP-TH 0110026;%%

%\cite{Salmhofer:2001tr}
\bibitem{Salmhofer:2001tr}
  M.~Salmhofer and C.~Honerkamp,
%   ``Fermionic renormalization group flows: Technique and theory,''
  Prog.\ Theor.\ Phys.\  {\bf 105} (2001) 1.
  %%CITATION = PTPKA,105,1;%%


%\cite{Litim:2000ci}
\bibitem{Litim:2000ci}
  D.~F.~Litim,
  %``Optimisation of the exact renormalisation group,''
  Phys.\ Lett.\ B {\bf 486} (2000) 92
  [hep-th/0005245].
  %%CITATION = HEP-TH 0005245;%%

%\cite{Litim:2001up}
\bibitem{Litim:2001up}
  D.~F.~Litim,
  %``Optimised renormalisation group flows,''
  Phys.\ Rev.\ D {\bf 64} (2001) 105007
  [hep-th/0103195].
  %%CITATION = HEP-TH 0103195;%%


%\cite{Litim:2001fd}
\bibitem{Litim:2001fd}
  D.~F.~Litim,
  %``Mind the gap,''
  Int.\ J.\ Mod.\ Phys.\ A {\bf 16} (2001) 2081
  [hep-th/0104221].
  %%CITATION = HEP-TH 0104221;%%

%\cite{Pawlowski:2005xe}
\bibitem{Pawlowski:2005xe}
  J.~M.~Pawlowski,
  %``Aspects of the functional renormalisation group,''
  hep-th/0512261.
  %%CITATION = HEP-TH 0512261;%%




%\cite{Dyson:1968up}
\bibitem{Dyson:1968up}
  F.~J.~Dyson,
  %``Existence Of A Phase Transition In A One-Dimensional Ising Ferromagnet,''
  Commun.\ Math.\ Phys.\  {\bf 12} (1969) 91.
  %%CITATION = CMPHA,12,91;%%

%\cite{Wilson:1971dh}
\bibitem{Wilson:1971dh}
  K.~G.~Wilson,
  %``Renormalization group and critical phenomena. 2. Phase space cell analysis
  %of critical behavior,''
  Phys.\ Rev.\  B {\bf 4} (1971) 3184,
  %%CITATION = PHRVA,B4,3184;%%
%\cite{Wilson:1972}
%\bibitem{Wilson:1972}
%  K.~G.~Wilson,
  Phys.\ Rev.\  D {\bf 6} (1972) 419.

%\cite{Baker:1972}
\bibitem{Baker:1972} G.~Baker, Phys.~Rev.~B~{\bf 5} (1972) 2622.

%\cite{Golner:1973}
\bibitem{Golner:1973}
G.~Golner, Phys.~Rev.~B~{\bf 8} (1973) 339.

%\cite{Meurice:2007zg}
\bibitem{Meurice:2007zg}
  Y.~Meurice,
  %``Global aspects of the renormalization group flows of Dyson's hierarchical
  %model,''
  hep-th/0701191.
  %%CITATION = HEP-TH/0701191;%%


%\cite{KochWittwer:1988}
\bibitem{KochWittwer:1988} H.~Koch and P.~Wittwer, in: {\it Nonlinear
    Evolution and Critical Phenomena}, NATO Advanced Study Institute,
  Vol.~176, Eds.~G.~Gallavotti and P.~Zweifel, Plenum Press (1988), p.~269.



%\cite{Pinn:1994st}
\bibitem{Pinn:1994st}
  K.~Pinn, A.~Pordt and C.~Wieczerkowski,
  %``Algebraic Computation Of Hierarchical Renormalization Group Fixed Points
  %And Their Epsilon Expansions,''
  J.\ Statist.\ Phys.\  {\bf 77} (1994) 977
  [hep-lat/9402020];\\
  %%CITATION = HEP-LAT 9402020;%%
%\cite{Godina:1997dk}
%\bibitem{Godina:1997dk}
  J.~J.~Godina, Y.~Meurice, M.~B.~Oktay and S.~Niermann,
  %``A guide to precision calculations in Dyson's hierarchical scalar field
  %theory,''
  Phys.\ Rev.\ D {\bf 57} (1998) 6326
  [hep-lat/9709097];\\
  %%CITATION = HEP-LAT 9709097;%%
 %\cite{Godina:1998nh}
%\bibitem{Godina:1998nh}
  J.~J.~Godina, Y.~Meurice and M.~B.~Oktay,
  %``High accuracy calculations of the critical exponents of Dyson's
  %hierarchical model,''
  Phys.\ Rev.\ D {\bf 59} (1999) 096002
  [hep-lat/9810034];\\
  %%CITATION = HEP-LAT 9810034;%%
%\cite{Gottker-Schnetmann:1999eg}
%\bibitem{Gottker-Schnetmann:1999eg}
  J.~Gottker-Schnetmann,
  %``O(N)-invariant hierarchical renormalization group fixed points by
  %algebraic numerical computation and epsilon expansion,''
  cond-mat/9909418.
  %%CITATION = COND-MAT 9909418;%%


%\cite{Godina:1998uz}
\bibitem{Godina:1998uz}
  J.~J.~Godina, Y.~Meurice and M.~B.~Oktay,
  %``Accurate checks of universality for Dyson's hierarchical model,''
  Phys.\ Rev.\ D {\bf 57} (1998) 6581
  [hep-lat/9802001].
  %%CITATION = HEP-LAT 9802001;%%


%\cite{Polchinski:1983gv}
\bibitem{Polchinski:1983gv}
  J.~Polchinski,
  %``Renormalization And Effective Lagrangians,''
  Nucl.\ Phys.\ B {\bf 231} (1984) 269.
  %%CITATION = NUPHA,B231,269;%%

%\cite{Felder:1987}
\bibitem{Felder:1987}
  G.~Felder
  Commun.\ Math.\ Phys.\  {\bf 111} (1987) 101.
  %%CITATION = CMPHA,111,101;%%



%\cite{Litim:2005us}
\bibitem{Litim:2005us}
  D.~F.~Litim,
  %``Universality and the renormalisation group,''
  JHEP {\bf 0507} (2005) 005
  [hep-th/0503096].
  %%CITATION = HEP-TH 0503096;%%

%\cite{Morris:2005ck}
\bibitem{Morris:2005ck}
  T.~R.~Morris,
  %``Equivalence of local potential approximations,''
  JHEP {\bf 0507} (2005) 027
  [hep-th/0503161].
  %%CITATION = HEP-TH 0503161;%%



\bibitem{continuum}
C.\,Wetterich, Phys.~Lett.~B~{\bf 301} (1993) 90.


%\cite{Bervillier:2007rc}
\bibitem{Bervillier:2007rc}
  C.~Bervillier, A.~J\"uttner and D.~F.~Litim,
  %``High-accuracy scaling exponents in the local potential approximation,''
  hep-th/0701172.
  %%CITATION = HEP-TH/0701172;%%

%\cite{Litim:2001dt}
\bibitem{Litim:2001dt}
  D.~F.~Litim,
  %``Derivative expansion and renormalisation group flows,''
  JHEP {\bf 0111} (2001) 059
  [hep-th/0111159].
  %%CITATION = HEP-TH 0111159;%%


%\cite{Litim:2002cf}
\bibitem{Litim:2002cf}
  D.~F.~Litim,
  %``Critical exponents from optimised renormalisation group flows,''
  Nucl.\ Phys.\ B {\bf 631} (2002) 128
  [hep-th/0203006].
  %%CITATION = HEP-TH 0203006;%%


%\cite{Pawlowski:2003hq}
\bibitem{Pawlowski:2003hq}
  J.~M.~Pawlowski, D.~F.~Litim, S.~Nedelko and L.~von Smekal,
  %``Infrared behaviour and fixed points in Landau gauge QCD,''
  Phys.\ Rev.\ Lett.\  {\bf 93} (2004) 152002
  [hep-th/0312324];
  %%CITATION = PRLTA,93,152002;%%
%\cite{Pawlowski:2004ip}
%\bibitem{Pawlowski:2004ip}
%  J.~M.~Pawlowski, D.~F.~Litim, S.~Nedelko and L.~von Smekal,
  %``Signatures of confinement in Landau gauge QCD,''
  AIP Conf.\ Proc.\  {\bf 756} (2005) 278
  [hep-th/0412326];\\
  %%CITATION = APCPC,756,278;%%
%\cite{Litim:2004wx}
%\bibitem{Litim:2004wx}
  D.~F.~Litim, J.~M.~Pawlowski, S.~Nedelko and L.~V.~Smekal,
  %``Infrared QCD and the renormalisation group,''
  hep-th/0410241.
  %%CITATION = HEP-TH/0410241;%%

%\cite{Lauscher:2001rz}
\bibitem{Lauscher:2001rz}
  O.~Lauscher and M.~Reuter,
  %``Is quantum Einstein gravity nonperturbatively renormalizable?,''
  Class.\ Quant.\ Grav.\  {\bf 19} (2002) 483
  [hep-th/0110021];\\
  %%CITATION = CQGRD,19,483;%%
%\cite{Litim:2003vp}
%\bibitem{Litim:2003vp}
  D.~F.~Litim,
  %``Fixed points of quantum gravity,''
  Phys.\ Rev.\ Lett.\  {\bf 92} (2004) 201301
  [hep-th/0312114];\\
  %%CITATION = PRLTA,92,201301;%%
%\cite{Percacci:2005wu}
%\bibitem{Percacci:2005wu}
  R.~Percacci,
  %``Further evidence for a gravitational fixed point,''
  Phys.\ Rev.\  D {\bf 73} (2006) 041501
  [hep-th/0511177];\\
  %%CITATION = PHRVA,D73,041501;%%
%\cite{Fischer:2006fz}
%\bibitem{Fischer:2006fz}
  P.~Fischer and D.~F.~Litim,
  %``Fixed points of quantum gravity in extra dimensions,''
  Phys.\ Lett.\  B {\bf 638} (2006) 497
  [hep-th/0602203];  
%%CITATION = PHLTA,B638,497;%%
%\cite{Fischer:2006at}
%\bibitem{Fischer:2006at}
%  P.~Fischer and D.~F.~Litim,
  %``Fixed points of quantum gravity in higher dimensions,''
  AIP Conf.\ Proc.\  {\bf 861} (2006) 336
  [hep-th/0606135];\\
  %%CITATION = APCPC,861,336;%%
%
%\cite{Codello:2006in}
%\bibitem{Codello:2006in}
  A.~Codello and R.~Percacci,
  %``Fixed points of higher derivative gravity,''
  Phys.\ Rev.\ Lett.\  {\bf 97} (2006) 221301
  [hep-th/0607128].
  %%CITATION = PRLTA,97,221301;%%



%\cite{Litim:2006ag}
\bibitem{Litim:2006ag}
  D.~F.~Litim and J.~M.~Pawlowski,
  %``Non-perturbative thermal flows and resummations,''
  JHEP {\bf 0611} (2006) 026
  [hep-th/0609122];
  %%CITATION = JHEPA,0611,026;%%
%%\cite{Blaizot:2006rj}
%\bibitem{Blaizot:2006rj}
  J.~P.~Blaizot, A.~Ipp, R.~Mendez-Galain and N.~Wschebor,
  %``Perturbation theory and non-perturbative renormalization flow in scalar
  %field theory at finite temperature,''
  Nucl.\ Phys.\  A {\bf 784} (2007) 376
  [hep-ph/0610004].
  %%CITATION = NUPHA,A784,376;%%

%\cite{Blaizot:2004qa}
\bibitem{Blaizot:2004qa}
  J.~P.~Blaizot, R.~Mendez Galain and N.~Wschebor,
  %``Non Perturbative Renormalization Group, momentum dependence of $n$-point
  %functions and the transition temperature of the weakly interacting Bose
  %gas,''
  Europhys.\ Lett.\  {\bf 72} (2005) 705
  [cond-mat/0412481].
  %%CITATION = EULEE,72,705;%%

%\cite{Litim:1995ex}
\bibitem{Litim:1995ex}
  D.~F.~Litim and N.~Tetradis,
  %``Approximate solutions of exact renormalization group equations,''
  hep-th/9501042.
  %%CITATION = HEP-TH 9501042;%%



%\cite{Reuter:1993kw}
\bibitem{Reuter:1993kw}
  M.~Reuter and C.~Wetterich,
  %``Effective average action for gauge theories and exact evolution
  %equations,''
  Nucl.\ Phys.\  B {\bf 417} (1994) 181;\\
  %%CITATION = NUPHA,B417,181;%%
%\cite{Freire:2000bq}
%\bibitem{Freire:2000bq}
  F.~Freire, D.~F.~Litim and J.~M.~Pawlowski,
  %``Gauge invariance and background field formalism in the exact
  %renormalisation group,''
  Phys.\ Lett.\  B {\bf 495} (2000) 256
  [hep-th/0009110].%;\\
  %%CITATION = PHLTA,B495,256;%%

%\cite{Reuter:1997gx}
\bibitem{Reuter:1997gx}
  M.~Reuter and C.~Wetterich,
  %``Gluon condensation in nonperturbative flow equations,''
  Phys.\ Rev.\  D {\bf 56} (1997) 7893
  [hep-th/9708051];\\
  %%CITATION = PHRVA,D56,7893;%%
%\cite{Gies:2002af}
%\bibitem{Gies:2002af}
  H.~Gies,
  %``Running coupling in Yang-Mills theory: A flow equation study,''
  Phys.\ Rev.\  D {\bf 66} (2002) 025006
  [hep-th/0202207].
  %%CITATION = PHRVA,D66,025006;%%

%\cite{Litim:2002ce}
\bibitem{Litim:2002ce}
  D.~F.~Litim and J.~M.~Pawlowski,
  %``Renormalisation group flows for gauge theories in axial gauges,''
  JHEP {\bf 0209} (2002) 049
  [hep-th/0203005].
  %%CITATION = JHEPA,0209,049;%%

%\cite{Litim:2002hj}
\bibitem{Litim:2002hj}
  D.~F.~Litim and J.~M.~Pawlowski,
  %``Wilsonian flows and background fields,''
  Phys.\ Lett.\  B {\bf 546} (2002) 279
  [hep-th/0208216],
  %%CITATION = PHLTA,B546,279;%%
%\cite{Litim:2001hk}
%\bibitem{Litim:2001hk}
%  D.~F.~Litim and J.~M.~Pawlowski,
  %``Predictive power of renormalisation group flows: A comparison,''
  Phys.\ Lett.\  B {\bf 516} (2001) 197
  [hep-th/0107020].
  %%CITATION = PHLTA,B516,197;%%



%\cite{Litim:2002xm}
\bibitem{Litim:2002xm}
  D.~F.~Litim and J.~M.~Pawlowski,
  %``Completeness and consistency of renormalisation group flows,''
  Phys.\ Rev.\  D {\bf 66} (2002) 025030
  [hep-th/0202188],
  %%CITATION = PHRVA,D66,025030;%%
%\cite{Litim:2001ky}
%\bibitem{Litim:2001ky}
%  D.~F.~Litim and J.~M.~Pawlowski,
  %``Perturbation theory and renormalisation group equations,''
  Phys.\ Rev.\  D {\bf 65} (2002) 081701
  [hep-th/0111191].
  %%CITATION = PHRVA,D65,081701;%%

%\cite{Liao:1997nm}
\bibitem{Liao:1997nm}
S.~B.~Liao,
%``Operator Cutoff Regularization and Renormalization Group in Yang-Mills Theory,''
Phys.\ Rev.\  {\bf D56} (1997) 5008
[hep-th/9511046];
%%CITATION = HEP-TH 9511046;%%
%\cite{Liao:1996fp}
%\bibitem{Liao:1996fp}
%S.~B.~Liao,
%``On connection between momentum cutoff and the proper time regularizations,''
Phys.\ Rev.\  {\bf D53} (1996) 2020.
%[hep-th/9501124].
%%CITATION = HEP-TH 9501124;%%


%\cite{Bonanno:2000yp}
\bibitem{Bonanno:2000yp}
  A.~Bonanno and D.~Zappala,
  %``Towards an accurate determination of the critical exponents with the
  %renormalization group flow equations,''
  Phys.\ Lett.\  B {\bf 504} (2001) 181
  [hep-th/0010095];\\
  %%CITATION = PHLTA,B504,181;%%
%\cite{Mazza:2001bp}
%\bibitem{Mazza:2001bp}
  M.~Mazza and D.~Zappala,
  %``Proper time regulator and renormalization group flow,''
  Phys.\ Rev.\  D {\bf 64} (2001) 105013
  [hep-th/0106230].
  %%CITATION = PHRVA,D64,105013;%%

%\cite{Schaefer:1999em}
\bibitem{Schaefer:1999em}
  B.~J.~Schaefer and H.~J.~Pirner,
  %``The equation of state of quarks and mesons in a renormalization group  flow
  %picture,''
  Nucl.\ Phys.\  A {\bf 660} (1999) 439
  [nucl-th/9903003];\\
  %%CITATION = NUPHA,A660,439;%%
%\cite{Schaefer:2006sr}
%\bibitem{Schaefer:2006sr}
  B.~J.~Schaefer and J.~Wambach,
  %``Renormalization group approach towards the QCD phase diagram,''
  hep-ph/0611191;
  %%CITATION = HEP-PH/0611191;%%
%\cite{Schaefer:2006ds}
%\bibitem{Schaefer:2006ds}
%  B.~J.~Schaefer and J.~Wambach,
  %``Susceptibilities near the QCD (tri)critical point,''
  hep-ph/0603256.
  %%CITATION = HEP-PH/0603256;%%



 

%\cite{Meurice:1996bh}
\bibitem{Meurice:1996bh}
  Y.~Meurice and G.~Ordaz,
  %``A two-parameter recursion formula for scalar field theory,''
  J.\ Phys.\ A {\bf 29} (1996) L635
  [hep-lat/9608023].
  %%CITATION = HEP-LAT 9608023;%%




%\cite{Ball:1995ji}
\bibitem{Ball:1995ji}
R.~D.~Ball, P.~E.~Haagensen, J.~I.~Latorre and E.~Moreno,
%``Scheme independence and the exact renormalisation group,''
Phys.\ Lett.\  {\bf B347} (1995) 80;\\% [hep-th/9411122];\\
%%CITATION = HEP-TH 9411122;%%
%\href{\wwwspires?eprint=HEP-TH/9411122}{SPIRES}
%\cite{Litim:1997nw}
%\bibitem{Litim:1997nw}
D.~F.~Litim,
%``Scheme independence at first order phase transitions 
%and the  renormalisation group,''
Phys.\ Lett.\  {\bf B393} (1997) 103
[hep-th/9609040];\\
%%CITATION = HEP-TH 9609040;%%
%\cite{Liao:1999sh}
%\bibitem{Liao:1999sh}
  S.~B.~Liao, J.~Polonyi and M.~Strickland,
  %``Optimization of renormalization group flow,''
  Nucl.\ Phys.\ B {\bf 567} (2000) 493
  [hep-th/9905206];\\
  %%CITATION = HEP-TH 9905206;%%
%\cite{Freire:2001sx}
%\bibitem{Freire:2001sx}
F.~Freire and D.~F.~Litim,
%``Charge cross-over at the U(1)-Higgs phase transition,''
Phys.\ Rev.\ D {\bf 64} (2001) 045014
[hep-ph/0002153];\\
%%CITATION = HEP-PH 0002153;%%
%\cite{Canet:2003qd}
%\bibitem{Canet:2003qd}
  L.~Canet, B.~Delamotte, D.~Mouhanna and J.~Vidal,
  %``Nonperturbative renormalization group approach to the Ising model: a
  %derivative expansion at order $\partial^4$,''
  Phys.\ Rev.\  B {\bf 68} (2003) 064421
  [hep-th/0302227].
  %%CITATION = PHRVA,B68,064421;%%


%\cite{Litim:2003kf}
\bibitem{Litim:2003kf}
  D.~F.~Litim and L.~Vergara,
  %``Subleading critical exponents from the renormalisation group,''
  Phys.\ Lett.\  B {\bf 581} (2004) 263
  [hep-th/0310101].
  %%CITATION = PHLTA,B581,263;%%


\end{thebibliography}
\end{document}